# Four-Dimensional Higher-Order Chern Insulator and Its Acoustic Realization

Ze-Guo Chen[1], Weiwei Zhu[1,2], Yang Tan[1,3], Licheng Wang[1,4], and Guancong Ma[1,*]

[1]Department of Physics, Hong Kong Baptist University, Kowloon Tong, Hong Kong

We present a theoretical study and experimental realization of a system that is simultaneously a four-dimensional (4D) Chern insulator and a higher-order topological insulator (HOTI). The system sustains the coexistence of (*4-1*)-dimensional chiral topological hypersurface modes (THMs) and (*4-2*)-dimensional chiral topological surface modes (TSMs). Our study reveals that the THMs are protected by second Chern numbers, and the TSMs are protected by a topological invariant composed of two first Chern numbers, each belonging a Chern insulator existing in sub-dimensions. With the synthetic coordinates fixed, the THMs and TSMs respectively manifest as topological edge modes (TEMs) and topological corner modes (TCMs) in the real space, which are experimentally observed in a 2D acoustic lattice. These TCMs are not related to quantized polarizations, making them fundamentally distinctive from existing examples. We further show that our 4D topological system offers an effective way for the manipulation of the frequency, location, and the number of the TCMs, which is highly desirable for applications.

## I. INTRODUCTION

Topological phase is an important development and unexplored freedom of traditional band theories [1,2]. The universality of topological phases is exemplified in a wide variety of systems, such as solid-state electronic systems [1,2], photonics [3,4], cold atoms [5], acoustics and mechanics [6,7]. Recent studies have revealed a new class of "higher-order topological insulators" (HOTIs), which refer to a *d*-dimensional topologically nontrivial system that can sustain (*d-n*)-dimensional boundary modes, with *n* > 1 [8-23]. For example, 0D topological corner modes (TCMs) can be found in 2D systems. Although the studies of HOTIs have led to several significant developments, these second-order TCMs generally do not coexist with first-order topologically protected gapless edge modes [11].

On the other hand, topological phases can also arise in parameter space that is spanned by both spatial (or equivalently, reciprocal) and synthetic dimensions [24-30]. A notable example is the realization of the Hofstadter butterfly, which was originally proposed in a two-dimensional (2D) square lattice, in a 1D system by introducing one additional synthetic dimension [31]. Weyl points, which are widely studied in 3D periodic systems, have also been demonstrated using a system with one spatial and two synthetic dimensions [32,33]. Synthetic dimensions also enable the investigation of systems that go beyond 3D, with the 4D quantum Hall effect being an important example [25,34,35]. The smart use of the extra dimensionality has led to an exciting array of novel phenomena such as quantum Hall effect in quasicrystals [25] and topological charge pumping [29,34,35]. However, so far the higher-order topological modes in 4D systems remain unexplored and have not been realized.

In this work, we study a 4D topological system consisting of two spatial and two synthetic dimensions. We find that the system is simultaneously a 4D Chern insulator [36] and a 4D HOTI. The system is gapless when truncated in the real space, in which case both ($4-1$)-dimensional chiral topological hypersurface modes (THMs) and ($4-2$)-dimensional second-order chiral topological surface modes (TSMs) coexist. THMs are protected by the second Chern numbers of the 4D bulk bands and TSMs are protected by nonzero combinations of first Chern numbers each belonging to a Chern insulator existing on orthogonal sub-dimensions. When both synthetic coordinates are fixed, the 4D system is observable as 2D real-space systems, wherein the THMs become 1D topological edge modes (TEMs) and the TSMs manifest as 0D topological corner modes (TCMs). Our findings are experimentally validated using a 2D acoustic lattice. Notably, due to their 4D topological origin, the TEMs and TCMs in real-space systems are fundamentally different from previously reported cases [13-22]. On the other hand, we identify that the THMs and TSMs can be mathematically traced to the topological boundary modes of the 2D Chern insulators [25,34,35]. This new perspective leads to striking capability for realizing TCMs and for manipulating their frequencies, locations, and number. Such capability is experimentally demonstrated by the realization of two distinctive types of TCMs, one is a "separable bound state in a continuum (BIC)" [37], and the other is the realization of multiple TCMs in one corner.

## II. A 4D CHERN INSULATOR REALIZED WITH TWO SYNTHETIC DIMENSIONS

First, we develop the theoretical model of a 4D Chern insulator and analyze its topological characteristics. Our 4D system consists of two spatial (or reciprocal) and two synthetic dimensions. To best introduce the idea, we begin by demonstrating a 2D Chern insulator with one spatial and one synthetic dimension. Consider a 1D chain of identical atoms in the $x$-direction, each coupled to its nearest neighbor through hopping $t$. The atomic chain is described by a tight-binding model

$$\widehat{H} = \sum_m (f_m |m\rangle\langle m| + t|m\rangle\langle m+1| + t|m+1\rangle\langle m|), \tag{1}$$

where $|m\rangle$ is the Dirac ket for site-$m$, $t$ is the hopping constant. We enforce a modulation to the onsite eigenfrequency

$$f_m(\phi_x) = f_0 + \lambda_x \cos(2\pi m b_x + \phi_x), \tag{2}$$

where $\lambda_x$ is the amplitude of onsite potential, and $b_x$ is the modulation frequency. The modulation has a phase factor $\phi_x$, which can be regarded as a pseudo-momentum that constitutes a synthetic dimension in our system, as shown in Fig. 1(a). Here, we set $b_x = p/q = 1/3$, making the system a commensurate one. We investigate a finite chain with 32 sites. The parameters used in the tight-binding models are $f_0 = 2095$ Hz, $t = -124.75$ Hz, $\lambda/t = -1.9$, which are related to the acoustic system which will be discussed. The procedures for determining these parameters are presented in ref. [38]. The Hamiltonian satisfies $\widehat{H}(\phi_x) = P^{-1}\widehat{H}(-\phi_x)P$, where the nonzero element of the unitary operator $P$ is defined as $P_{ij} = 1$ for $i + j = N + 1$. This indicates the band structure is symmetric about $\phi_x = 0$, as shown in Fig. 1(b). We have computed the Chern numbers $C_G$ in the $k_x \phi_x$-plane for the 1st and 2nd bulk bandgap and both gaps have nonzero Chern numbers, which confirms that the system is a 2D Chern insulator. As a result, the system is gapless, and two chiral gapless boundary modes are clearly identified (Fig. 1(b, c)). It is noteworthy that this Chern insulator only involves modulation to onsite energy, whereas hopping $t$ remains constant. This important characteristic sets our system apart from the widely used the Su-Schrieffer-Heeger model, which relies on staggered hopping but has identical onsite energy.

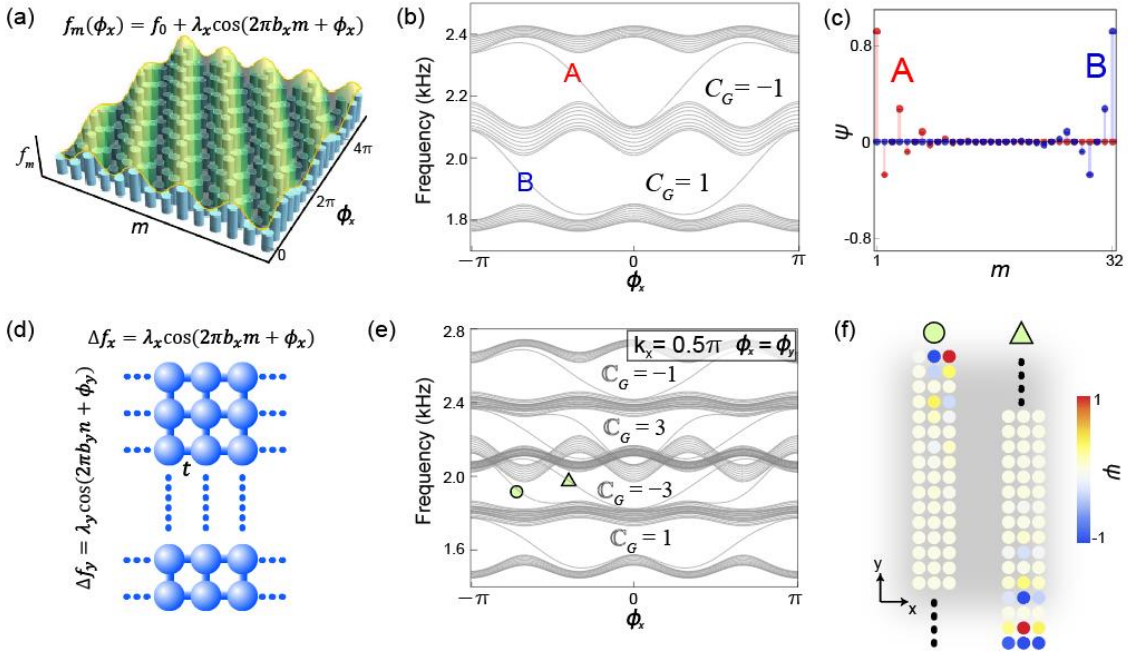

**FIG. 1 4D Chern insulator.** (a) A schematic drawing of a 2D Chern insulator with one spatial and one synthetic dimension. The onsite frequency is modulated as $f_m(\phi_x)$, with $\phi_x$ becoming the synthetic dimension. (b) The eigenfrequencies as functions of $\phi_x$ calculated using the tight-binding model (Eqs. (1, 2)). Nonzero gap Chern numbers are marked in the bandgaps. Two chiral boundary modes are shown in (c). A 4D Chern insulator can be attained using two spatial and two synthetic dimensions. Using a ribbon which is periodic in $x$ but finite in $y$ with $N_y = 32$ sites (d), we can compute the system's eigenspectra. (e) shows the eigenspectra as a function of $\phi_x$ sliced at $k_x = 0.5\pi$ and $\phi_x = \phi_y$. The bandgaps are associated with nonzero second Chern numbers as labeled. The bulk bands are closed by THMs localized at the $x$-direction edges. The real-space distributions of two THMs examples are shown in (f). All results here were obtained using a tight-binding model.

By incorporating two synthetic dimensions, a 4D Chern insulator can be constructed using a square lattice of nearest-coupled sites, with each onsite frequency $f_0$ is modulated to

$$f_{m,n}(\phi_x, \phi_y) = f_0 + \lambda_x \cos(2\pi b_x m + \phi_x) + \lambda_y \cos(2\pi b_y n + \phi_y), \quad (3)$$

where, $m$, $n$ label the sites, $b_x$ and $b_y$ are the modulation frequencies in the $x$ and $y$ direction, and $\phi_x$, $\phi_y$ are the respective modulation phase factors. The system can be described by a tight-binding Hamiltonian

$$\widehat{\mathbb{H}}(\phi_x, \phi_y) = \sum_{m,n} \begin{bmatrix} f_{m,n}(\phi_x, \phi_y)|m,n\rangle\langle m,n| \\ +(t|m,n\rangle\langle m+1,n| + t|m,n\rangle\langle m,n+1|) + \text{h.c.} \end{bmatrix}, \quad (4)$$

where $|m,n\rangle$ is the Dirac ket for site $(m,n)$. Note that Eq. (3) implies that the modulation in the $x$ and $y$ directions are independent, consequently $\phi_x$ and $\phi_y$ constitute two orthogonal dimensions. Hence Eq. (4) describes a system living in a 4D space spanned by $(k_x, k_y, \phi_x, \phi_y)$. It also suggests that $b_x, \lambda_x, b_y, \lambda_y$ can be independently tuned, which we will later explore. Here, we set the modulation frequencies to be $b_x = b_y = 1/3$ and the modulation amplitudes $\lambda_x = \lambda_y = -1.9t$. A unit cell contains $(b_x b_y)^{-1} = 9$ sites therefore the system has nine bulk bands. We find that these bands form five bulk band regions separated by four bandgaps. The nontrivial topology of the 4D system is characterized by the second Chern number for bulk bands [36,39]

$$\mathbb{C}_B = \frac{1}{32\pi^2} \int d^4\phi \, \varepsilon_{ijkl} \, \text{Tr}\left[F_{ij}^{\alpha\beta} F_{kl}^{\alpha\beta}\right], \quad (5)$$

with $F_{ij}^{\alpha\beta} = \partial_i A_j^{\alpha\beta} - \partial_j A_i^{\alpha\beta} + i[A_i, A_j]^{\alpha\beta}$ and $A_i^{\alpha\beta}(\boldsymbol{\phi}) = -i\left\langle \alpha, \boldsymbol{\phi} \middle| \frac{\partial}{\partial \phi_i} \middle| \beta, \boldsymbol{\phi} \right\rangle$. In Eq. (5), $\varepsilon_{ijkl}$ is an antisymmetric tensor of rank 4, $(i,j,k,l)$ index the four dimensions: $k_x, k_y$ and $\phi_x, \phi_y$. $F_{ij}^{\alpha\beta}$ is the 2D Berry curvature for a state defined in pseudo-momentum space $i,j$, with $\alpha, \beta$ referring to the occupied multiple bands. The second Chern numbers for 4D bandgaps, denoted $\mathbb{C}_G$, can be obtained by adding the $\mathbb{C}_B$ of all the bands below that gap. We find that $\mathbb{C}_G$ for the four bandgaps are 1, –3, 3, –1, respectively [40]. Our system is therefore a 4D Chern insulator.

From the bulk-surface correspondence, a nonzero second Chern number implies the existence of first-order $(4–1)$-dimensional chiral topological modes. To investigate, we employ a 4D "ribbon" that is periodic in $x$, $\phi_x$, $\phi_y$ but finite in $y$. We cut the eigenspectra at $k_x = 0.5\pi$ and along the line of $\phi_x = \phi_y$, the results are plotted as functions of $\phi_x$ shown in Fig. 1(e). It is seen that the system is indeed gapless, with its five well-defined bulk band regions connected by chiral boundary modes. In Fig. 1(f), we can see that the topological modes exponentially decay in the $y$-direction in the real space. In other words, they exist on the $k_x \phi_x \phi_y$-hyperplane. We therefore called them $(4–1)$-dimensional chiral topological hypersurface modes (THMs). Similarly, when the 4D system is truncated in $x$-directions, THMs are found on the $k_y \phi_x \phi_y$-hyperplane.

## III.  A 4D HIGHER-ORDER CHERN INSULATOR

Our system is also a 4D HOTI. To see this, we consider the same 4D system that is finite in both $x$ and $y$. Same as before, there are five bulk band regions separated by four bandgaps (Fig. 2(a)). Connecting these bulk bands are two sets of ($4$–$1$)-dimensional chiral THMs, sustained on the $k_x\phi_x\phi_y$- and $k_y\phi_x\phi_y$-hyperplanes, respectively. These are plotted in Fig. 2(b) and (c). Meanwhile, four ($4$–$2$)-dimensional second-order topological surface modes (TSMs) are identified (Fig. 2(d)). Note that the green surface actually contains two degenerate TSMs. The TSMs live entirely on the $\phi_x\phi_y$-plane, and exponentially decay in both $x$- and $y$-directions, as shown in Fig. 2(e). A striking observation is that these TSMs are dispersive in the two synthetic dimensions and exist entirely within the THM bandgaps, making the THMs gapless. Hence they are second-order chiral topological modes. Since the THMs are found in 4D bulk gaps, the TSMs can essentially overlap with the bulk bands in frequency, implying the existence of 2D bound states in a 4D continuum, which we will demonstrate in an experiment.

The topological nature of the TSMs can be revealed by considering the Hamiltonian of the finite-system (Eq. (4)). We observe that a finite system with $N \times N$ sites can be decomposed into two orthogonal copies of 2D Chern insulators (Eq. (1)). Mathematically, this is expressed as

$$\mathbb{H}(\phi_x, \phi_y) + f_0 I_{N^2} = I_N \otimes H_x(\phi_x) + H_y(\phi_y) \otimes I_N, \tag{6}$$

where $H_x(\phi_x)$ ($H_y(\phi_y)$) is the Hamiltonian of a 2D Chern system with $x(y)$ being the real dimension, $I_N$ ($I_{N^2}$) is an $N(N^2)$-dimensional identity matrix, and $\otimes$ denotes the Kronecker product. The right-hand side in the Eq. (6) introduces an additional onsite energy $f_0$ which is accounted for on the left-hand side. Eq. (6) reveals the mathematical separability of $\mathbb{H}(\phi_x, \phi_y)$, which implies that $H_x(\phi_x)$ and $H_y(\phi_y)$ exist on two orthogonal planes $k_x\phi_x$ and $k_y\phi_y$, yet these two planes do not meet. Such geometric orthogonality fundamentally roots in a 4D space. Physically, it indicates that the 4D Chern insulator can be decomposed to two independent copies of 2D Chern insulators. Eq. (6) also suggests that the eigenfunctions of $\mathbb{H}(\phi_x, \phi_y)$, denoted $|\Psi\rangle$, are given by the Kronecker product of the eigenfunctions from two 2D Chern system

$$|\Psi\rangle = |\psi_y\rangle \otimes |\psi_x\rangle. \tag{7}$$

wherein $|\psi_x\rangle$ and $|\psi_y\rangle$ are the eigenfunctions of $H_x(\phi_x)$ and $H_y(\phi_y)$, respectively. The eigenvalues of $\mathbb{H}(\phi_x, \phi_y) + f_0 I_{N^2}$ are given by the Minkowski sum of the eigenvalues of $H_x(\phi_x)$ and $H_y(\phi_x)$. These relations make our 4D system analytical, since the separated 2D Chern system can be analytically solved [41]. Detailed discussions about the formation rules of the 4D eigenmodes and eigenfrequencies are presented in [38]. It follows that the TSMs are composed by the chiral boundary modes of the 2D Chern insulators existing in the sub-dimensions, which are protected by nonzero gap Chern numbers $C_G^x$ and $C_G^y$, respectively. As a result, the TSMs are topologically protected by a non-zero topological invariant $\mathcal{C} \equiv (C_G^x, C_G^y)$.

Although the topological invariant $\mathcal{C}$ is seemly composed of two Chern numbers computed for 2D subsystems, it is fundamentally determined by the system's 4D topology. To see this, note that Eq. (6) implies that $F_{k_x\phi_x}^{\alpha\beta}$ and $F_{k_y\phi_y}^{\alpha\beta}$ are the only nonzero terms in Eq. (5). As a result, Eq. (5) is simplified to $\mathbb{C}_B = \frac{1}{2\pi}\int d^2\phi F_{k_x\phi_x} \times \frac{1}{2\pi}\int d^2\phi F_{k_y\phi_y}$, i. e., the product of two nonzero first Chern numbers [38,42]. Since the 4D bandgaps are well defined in our system, it is straightforward to consider the topology of bandgaps. The second Chern number of a 4D bandgap located near energy $\epsilon$ is related to the first Chern numbers of bands of 2D subsystems with energy $\epsilon_x + \epsilon_y < \epsilon$ [25],

$$\mathbb{C}_{G,\epsilon} = \sum_{\epsilon_x+\epsilon_y<\epsilon} C_{B,\epsilon_x}^x C_{B,\epsilon_y}^y. \tag{8}$$

Eq. (8) helps us to build a connection between the second Chern number and the topological invariant $\sum C_G^x C_G^y$ describing the number of TSMs, where the summation is only defined in the same gap of THMs. Specifically, in our system, Eq. (8) shows that the topological invariant protecting the first TSM, i. e., $\mathcal{C}_1 = (C_{G,1}^x, C_{G,1}^y) = (1,1)$, is related to the second Chern number for the first 4D bulk gap, $C_{G,1}^x C_{G,1}^y = C_{B,1}^x C_{B,1}^y = \mathbb{C}_{G,1} = 1$, in which the subscript numbers are the indices for bandgaps. For the second and third TSMs which are degenerate, we have $\mathcal{C}_2 = (C_{G,1}^x, C_{G,2}^y) = (1,-1)$, $\mathcal{C}_3 = (C_{G,2}^x, C_{G,1}^y) = (-1,1)$, there are $C_{G,1}^x C_{G,2}^y + C_{G,2}^x C_{G,1}^y = C_{B,1}^x C_{B,1}^y + (C_{B,1}^x C_{B,1}^y + C_{B,1}^x C_{B,2}^y + C_{B,2}^x C_{B,1}^y) = \mathbb{C}_{G,1} + \mathbb{C}_{G,2} = -2$. Likewise, for the fourth TSM, $\mathcal{C}_4 = (C_{G,2}^x, C_{G,2}^y) = (-1,-1)$ so that $C_{G,2}^x C_{G,2}^y = C_{B,1}^x C_{B,1}^y + C_{B,1}^x C_{B,2}^y + C_{B,2}^x C_{B,1}^y + C_{B,2}^x C_{B,2}^y = \mathbb{C}_{G,1} + \mathbb{C}_{G,2} - C_{B,1}^x C_{B,1}^y + (-C_{B,1}^x - C_{B,3}^x)(-C_{B,1}^y - $

$C_{B,3}^y) = \mathbb{C}_{G,1} + \mathbb{C}_{G,2} + \mathbb{C}_{G,3} = 1$.

It should be clear now that the TSMs in our system are conceptually distinctive from higher-order topological modes protected by nonzero quantized polarizations. The TSMs' topological protection is fundamentally tied to the 4D topological invariant. Such a relation between topological invariants in different dimensions is generally not present for quantized polarization. Discussion of the topological invariants for general cases with well-defined bandgaps is presented in [38].

We have also investigated the robustness of the THMs and TSMs against disorder. Notably, $\mathbb{C}_G$ remains unchanged as long as uncorrelated perturbations do not close the bulk gap, and THMs and TSMs both persist against these perturbations [25,38].

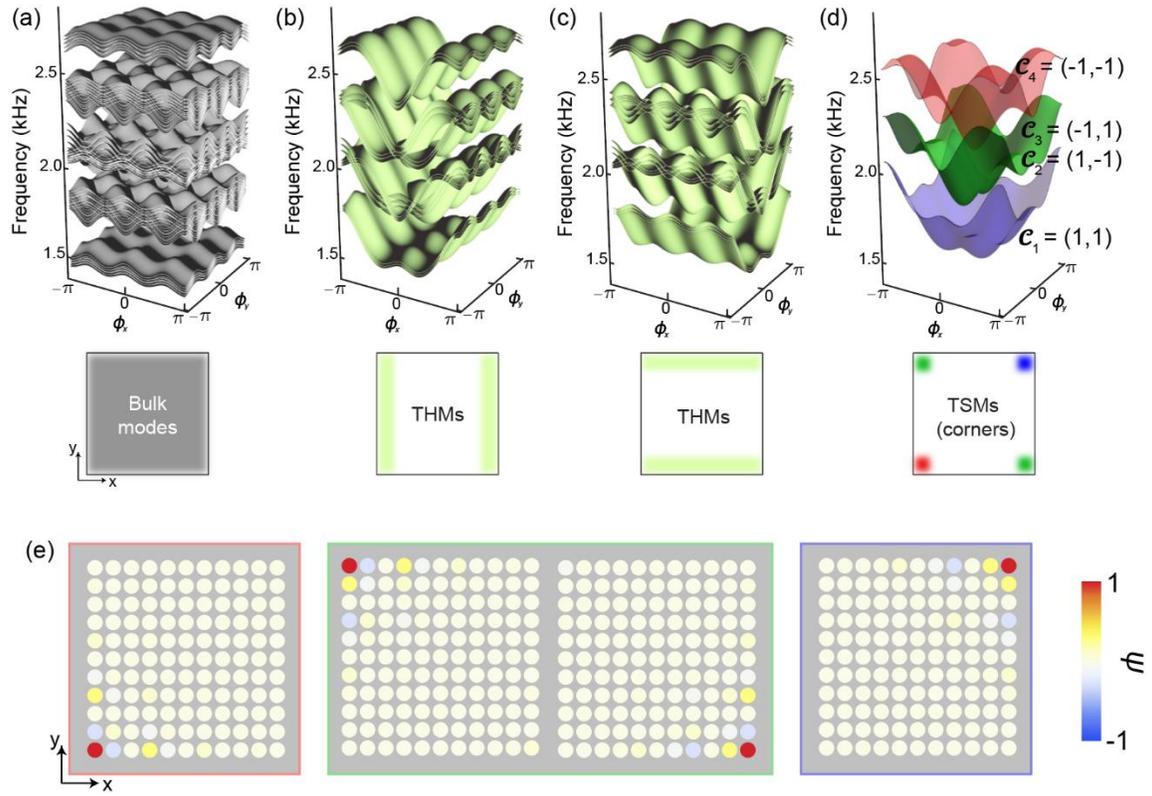

**FIG. 2 Second-order gapless TSMs in 4D Chern insulator** (a) 4D bulk modes occupy all four dimensions. Their eigenspectra are shown as functions of $(\phi_x, \phi_y)$, and they appear as 2D bulk modes in the real space (lower panel). (b, c) The THMs appear in the gaps of bulk bands and live on the $k_x\phi_x\phi_y$ and $k_y\phi_x\phi_y$-hyperplanes, respectively. They are localized on the edges in the real space. (d) The TSMs are found closing the THM gaps. The TSMs are 2D modes existing on $\phi_x\phi_y$-plane, therefore they are observed as 0D TCMs localized at the corners in the real space. The TSMs are colored to indicate their respective location. Note that the green sheets are two doubly degenerate states. (e) Real-space eigenfunctions of the four TSMs. The lattice here contains 11 ×

11 sites, which is the same as the acoustic system. All results here were obtained using a tight-binding model.

## IV. REALIZATION IN AN ACOUSTIC SYSTEM

Despite the THMs and TSMs are both protected by 4D topological invariants, they are observable in the 2D real-space systems once the synthetic coordinate of $\phi_x, \phi_y$ are fixed. As clearly shown in Fig. 1(f), the THMs emerge as topological edge modes (TEMs) in the real-space lattice. Meanwhile, the TSMs manifest as 0D TCMs localized at lattice corners (Fig. 2(e)).

We use a coupled acoustic cavity system, which is a proven platform for realizing tight-binding models [43], for the realization of our 4D system. We built a 2D acoustic lattice with 11×11 coupled cavities. The system is shown schematically in Fig. 3(a). All cavities have an initial height $h_0 = 120$ mm and a radius $r = 12$ mm. The cavities are sequentially connected at the top by a square tube with a side $d = 9$ mm. The outcome is a 2D periodic cavity lattice with a lattice constant $a = 40$ mm. The first longitudinal cavity mode, which has one node in the middle of the cavity (inset of Fig. 3(a)), is chosen as the onsite orbit. This mode's natural frequency is sensitive only to the height of the cavity. Therefore, the two synthetic dimensions $(\phi_x, \phi_y)$, which modulate the onsite eigenfrequencies, can be implemented by tuning the height of each cavity. We compute the eigenspectra of the 2D acoustic lattice using the commercial finite-element solver COMSOL Multiphysics (v5.4) along the parametric line $\phi_x = \phi_y$. The result is shown in Fig. 3(b) as functions of $\phi_x$, in which the THMs and TSMs are colored according to their real-space locations that are shown in the inset.

We note that some topological modes extend below the first band. This is due to the additional onsite perturbations caused by coupling tubes, which causes the eigenspectra to deviate from the ideal tight-binding model [38]. By accounting for this perturbation, we can reproduce the acoustic band structure using a modified tight-binding model with excellent agreement, as shown in Fig. 3(d).

Experimentally, the acoustic cavity system is machined from a block of aluminum and is filled with air. An aluminum plate was fixed on the block to seal the cavities and the coupling tubes. The top of each cavity has an opening port, which is used for excitation or measurement. The ports are blocked by plugs when not in use. For the measurement of the pressure response spectra, we used a waveform generator (Keysight 33500B) to send a short pulse covering 1,000 – 3,000 Hz to drive a loudspeaker that was placed on top of a chosen cavity. The response signals were received by a 1/4-inch microphone (PCB Piezotronics Model-378C10) and were then recorded by a digital oscilloscope (Keysight DSO2024A). The response spectra were then obtained by performing a Fourier transform on the transient signals. The measurements were repeated for each site to obtain the sound field distribution in the entire lattice. We then extract the data points at the frequencies of interest from the spectra. The results are normalized for each frequency. The two synthetic dimensions were implemented by injecting a specific volume of water into each cavity to adjust its height [44]. (The water surfaces are regarded as hard walls in the simulations due to the large impedance mismatch with air.) Three groups of parameters are experimentally adopted to demonstrate topological modes of different characteristics.

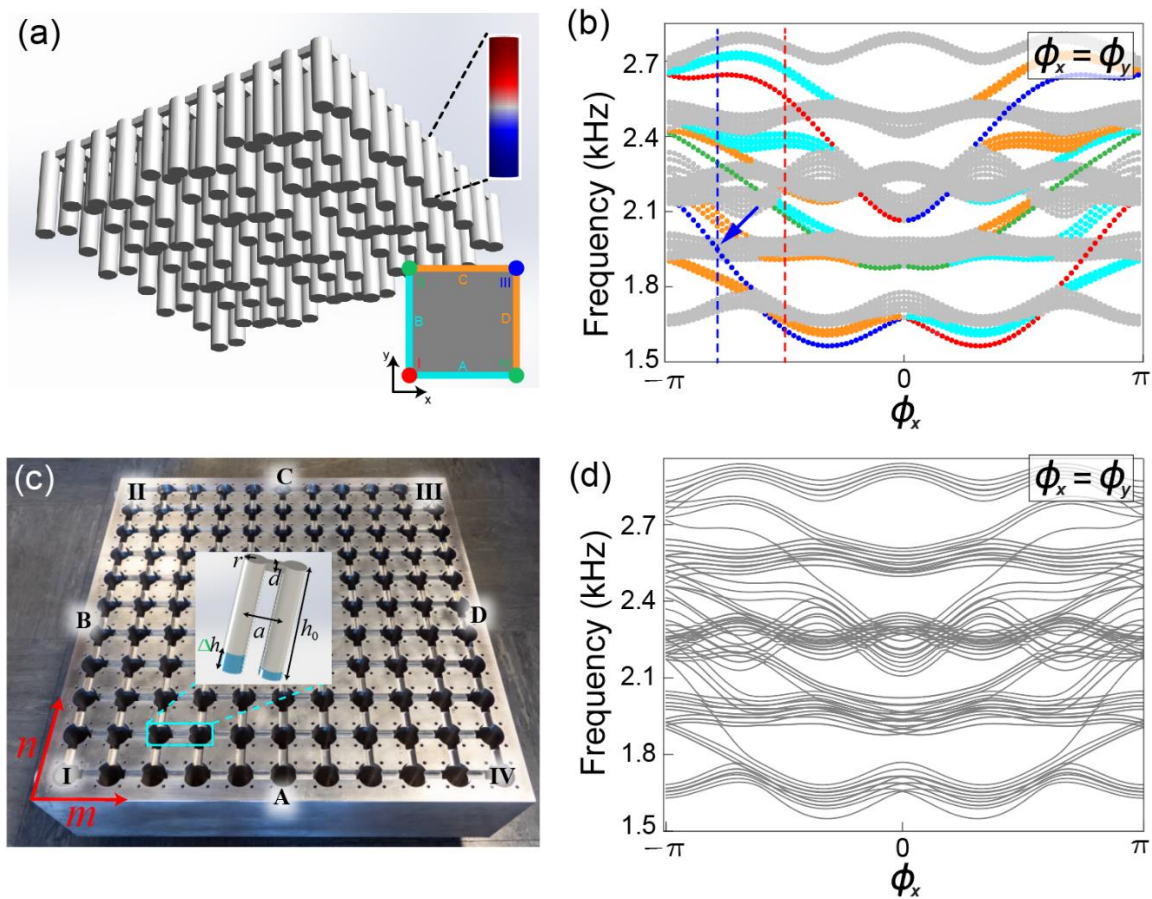

**FIG. 3 Realizing the 4D system using a 2D acoustic lattice.** (a) A schematic drawing of an acoustic system consisted of cavities with modulated heights. The inset shows the fundamental

cavity mode which is used as the onsite orbital, where red (blue) color represents positive (negative) sound pressure. In (d), the eigenspectra along $\phi_x = \phi_y$ are plotted as functions of $\phi_x$. The results are from finite-element simulations. The five gray regions are bulk bands, and the cyan /orange dots are in-gap THMs. The red/green/blue-colored dots indicate TSMs. The colors of modes indicate their characteristics, which are represented in the inset of (a). The blue arrow in (d) indicates a TSM that overlaps with bulk bands, making it a bound state in the continuum. The red dashed line marks $\phi_x = -0.5\pi$, the blue dashed line marks $\phi_x = -0.78\pi$. (c) A photograph of our acoustic lattice with 11×11 coupled cavities. To implement the two synthetic dimensions $(\phi_x, \phi_y)$, the height of each cavity is tuned by injecting a specific amount of water, as illustrated in the inset. (d) The eigenspectra along $\phi_x = \phi_y$ based on the modified tight-binding model. Excellent agreements with the results from simulations are seen.

### A. Observation of THMs and TSMs

First, we find that at $(\phi_x, \phi_y) = (-0.5\pi, -0.5\pi)$, which is marked by the red dashed line in Fig. 3(d), both THMs and TSMs can be observed in the 2D lattice. We tune the acoustic lattice to this point by precisely adding a specific amount of water into each cavity. The measured results are shown in Fig. 4. In Fig. 4(a), we schematically label the edges and corners using colors and tags. First, we drive the system with a loudspeaker at the center to excite the bulk modes. The measured response spectrum is shown in Fig 4(b) as a gray-shaded region. Five separate regions of high-pressure responses are clearly observed. We further raster-map the pressure response of all cavities at 2278 and 2538 Hz (marked by $f_1$, $f_2$), as shown in Fig. 4(c). These are extended modes in both spatial dimensions, clear evidence that they are the 4D bulk modes. Note that when $\phi_x = \phi_y$, the lattice possesses mirror symmetry along the line $x = y$ ($M_{x=y}$). This characteristic can be clearly identified in the field maps. Next, we identify that the system contains two sets of THMs, marked by cyan and orange to indicate their respectively real-space locations. As THMs are localized along one spatial dimension, we can observe them by exciting the acoustic system at the corresponding edges and measure their response spectra, which are shown in Fig. 4(b). Three peaks are seen for both edges A, B (cyan), and edges C, D (orange), which are consistent with our prediction as well as the simulation results in Fig. 3(d). Two cases of the spatial distributions of these modes are shown in Fig. 4(d), which clearly show that these are TEMs localized at the sample's edges, which agree well with our prediction. The TSMs, marked by red, blue, and green in Fig. 4(a), are 0D modes localized at the corners of the 2D lattice. By placing the source at the corresponding corner, we observe only one sharp resonant peak at each corner (Fig. 4(b)). Spatial pressure maps at each peak frequency further confirm that these modes are strongly localized at the corner and decay rapidly into the bulk (Fig. 4(e)). We note that the states at corners II and IV are ideally degenerate, owing to the system's mirror symmetry $M_{x=y}$ (along $x=y$). In the measured results, the two corresponding resonant peaks slightly mismatch in frequency (Fig. 4(b)). We attribute such discrepancy to experimental errors, which may cause $\phi_x, \phi_y$ to deviate from the ideal value. This is also a strong evidence that the existence of the THMs and TSMs are robust against disorders. In summary, the results confirm that the system possesses both TEMs and TCMs, which validates that the 4D system simultaneously supports both first-order THMs and second-order TSMs.

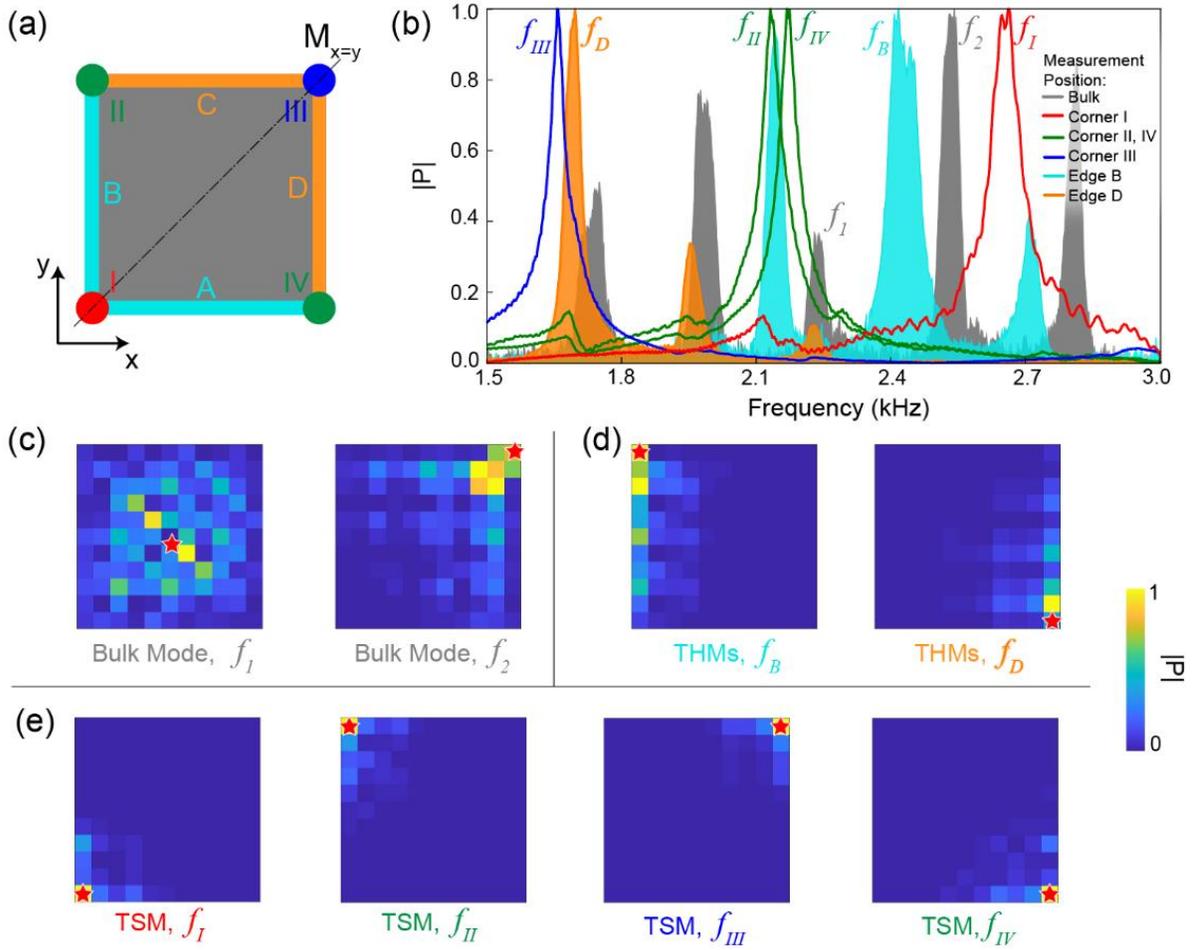

**FIG. 4. Observation of THMs and TSMs.** Here the system is at $(\phi_x, \phi_y) = (-0.5\pi, -0.5\pi)$. (a) A schematic drawing of the system, wherein the corners and edges are color-labeled. Note that the system has mirror symmetry $M_{x=y}$. (b) The pressure response spectra. The gray areas represent the bulk response; the orange/cyan-shaded areas are the edge responses; the four curves each represent the response at the correspondingly-colored corner. The spatial field maps are shown when the system is excited in the bulk (c), at the edges (d), and at the corners (e) at the indicated frequencies. The red stars in (c-e) mark the excitation position. In (d), the field maps are confined at the excitation edges, indicating the observation of THMs in real space as TEMs. In (e), the modes are strongly localized at the excitation corners, indicating the existence of TSMs in real space.

### B. TSM as a bound state in the continuum (BIC)

The fact that TSMs in our system are chiral modes closing the gaps of THMs has two implications. First, the TSMs are dispersive in the synthetic coordinates; second, as THMs are entirely in the 4D bulk gaps, the TSMs can overlap with the bulk bands in frequency, becoming bound states in the bulk continuum. An example can be seen near $(\phi_x, \phi_y) = (-0.78\pi, -0.78\pi)$, which is marked by the blue dashed line in Fig. 3(d). Since TSMs are TCMs in real space, they are observable as corner-mode BIC.

We tune the acoustic system to this parameter point by adjusting the amount water in each cavity. Fig. 5(a) shows the pressure response spectra of this case. When the excitation is at corner III, one single response peak is seen at $f_{BIC} = 2060$ Hz (red curve). This peak spectrally overlaps with the second bulk band (gray regions). We then place the source at corner III to excite at $f_{BIC}$ and obtain the field maps. A highly localized corner mode is clearly seen (Fig. 5(b)). In contrast, bulk modes are excited at the same frequency when the source is in the center (Fig. 5(c)) or at corner I (Fig. 5(d)). These results unambiguously show that the mode at corner III at $f_{BIC}$ is a BIC [37,45].

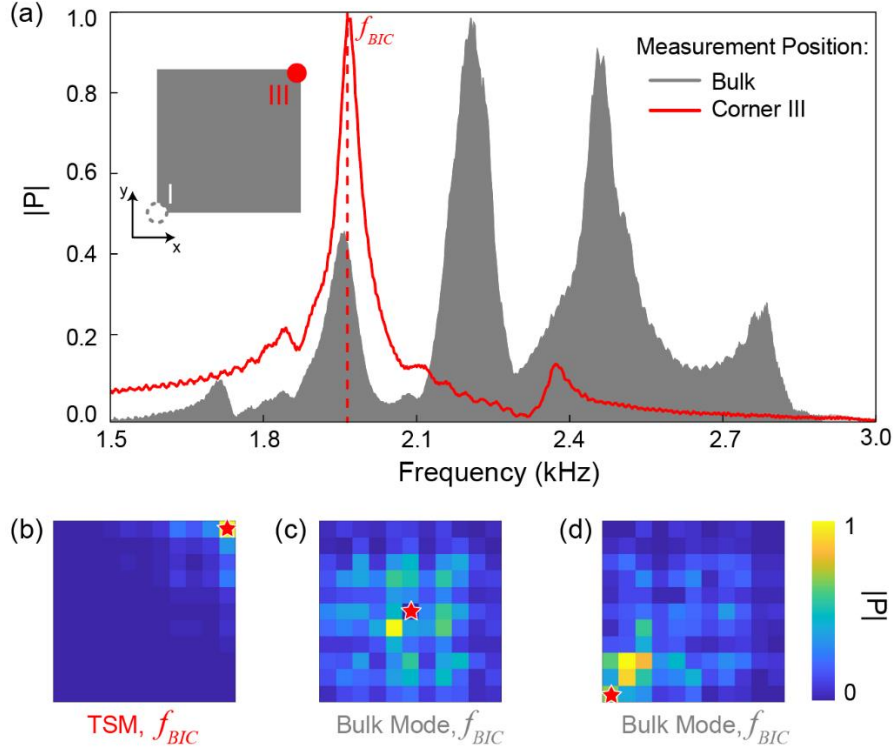

**FIG. 5. Using the gapless TSM to realize BIC in real space.** (a) The pressure responses measured at corner III (red) and in the bulk (gray region). A single peak at $f_{BIC} = 2060Hz$ is seen for the corner response, which overlaps with the second bulk band. (b) The spatial field map at $f_{BIC}$ when the source is at corner III. A highly localized corner mode is seen. (c) Excited at $f_{BIC}$ by a source at the center, extended modes occupying the entire bulk is seen. (d) In comparison, when the system is pumped at $f_{BIC}$ by a source placed at corner I, the field map indicates extended modes. The inset of (a) is a skematic drawing of the 2D acoustic lattice with the corners are color-tagged.

### C. Multiple TSMs localized at the same corner

The characteristics of THMs and TSMs are fully revealed only when considering all four dimensions, they are nevertheless observable as TEMs and TCMs in real space. This means that the real-space descendant system can also be regarded as a new type of 2D HOTI that simultaneously supports TEMs and TCMs. Moreover, the 4D system brings extra degrees of freedom in the manipulation of the TEMs and TCMs.

To show the unique advantage of our system, we set $b_x = 1/6$, $\lambda_x = -2t$ while keeping $b_y = 1/3$, $\lambda_y = -1.9t$. We analyze the point $(\phi_x, \phi_y) = (-0.6\pi, -0.28\pi)$ and find a total of five TSMs in this system, as shown in the eigenspectrum in Fig. 6(a). In the 2D lattice, three TSMs are localized at corner II and the other two localized at corner III. These are shown in Fig. 6(b). We validate these findings in our acoustic system. Our results show strong evidence for the existence of all five TSMs as corner modes. Three / two resonant peaks are clearly seen when the system is excited at corner II / III (Fig. 6(b)). The field distributions at the peak frequencies (Fig. 6(c)) indicate localized modes at their respective corners.

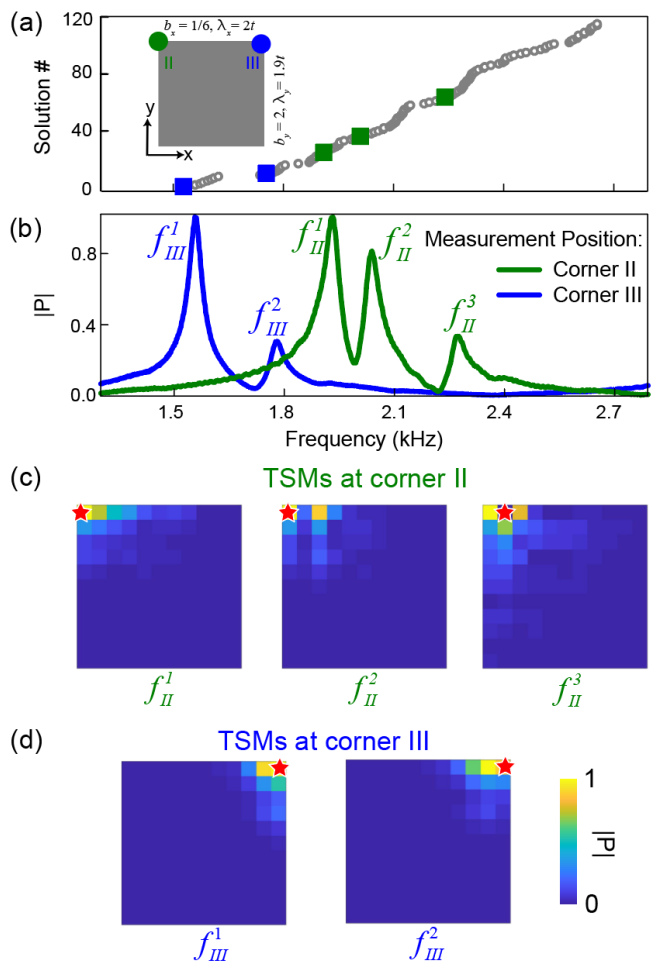

FIG. 6. **Exploiting the TSMs for the realization of multiple TCMs at the same corner.** (a) Eigenspectrum of a 4D acoustic system with $b_x = 1/6, b_y = 1/3$ at $(\phi_x, \phi_y) = (-0.6\pi, -0.28\pi)$. (b) The measured corner responses have three peaks for corner II (green) and two peaks for corner III (blue). The peak frequencies are consistent with the eigenfrequencies found in finite-element simulation. (c, d) are the field maps of the three TSMs at corner II and the two TSMs at corner III, respectively. The red stars mark the source position in each case. Note that the third TSM at corner II is excited with a source located one site away from the corner.

## II. DISCUSSION AND CONCLUSIONS

Our 4D topological system simultaneously sustains first-order THMs and second-order TSMs. Both the THMs and TSMs are gapless. In the real-space descendant system with the synthetic coordinates fixed, the TSMs become the TCMs to be observable in experiments. Hence these TCMs are fundamentally different from those reported previously, which are typically the consequence of non-zero corner charges induced by quantized polarizations. In contrast, the topological invariant protecting our TCMs are unveiled only by ascending to 4D. The topological invariant, which consists of two first Chern numbers, each responsible for a Chern insulator in a 2D plane, can only be meaningfully defined in a 4D hyperspace. Its 4D origin is further confirmed by its close tie to the second Chern numbers of the 4D gap, which we have shown in section III. On the other hand, despite the real-space system is a square or rectangular lattice, it does not possess any crystalline symmetry for most values of $\phi_{x,y}$. A direct consequence is that bulk polarizations are not quantized and cannot serve as a topological invariant in our system. This again fundamentally distinguishes our system from existing HOTIs.

The rich degrees of freedom offered by the 4D topology leads to a powerful recipe that brings unprecedented capabilities for tailoring higher-order topological modes. In particular, Eq. (7) suggests that all states of the 4D system can be composed by states in 2D subsystem. It follows that the eigenfrequencies of the 4D modes are given by the Minkowski sum of the 2D modes. We have further identified the 4D bulk modes' eigenfrequencies are given by $E_{4D}^{bulk} = E_x^{bulk} + E_y^{bulk} - f_0$, the THMs follow $E_{4D}^{THM} = E_x^{boundary} + E_y^{bulk} - f_0$ or $E_{4D}^{THM} = E_y^{boundary} + E_x^{bulk} - f_0$, and the TSMs follow $E_{4D}^{TSM} = E_x^{boundary} + E_y^{boundary} - f_0$. Here, $E_{x,y}^{bulk}, E_{x,y}^{boundary}$ are the eigenfrequencies of the bulk and boundary modes of the 2D subsystems described by $H_x(\phi_x)$ and $H_y(\phi_y)$, respectively. This consideration clearly allows the easy tracking and tuning of each 4D mode's eigenfrequency by independently considering the sub-dimensional systems, which are much easier to control. In addition, we further show in ref. [38] that this approach can lead to a flexible way to design the real-space location of THMs and TSMs, which offers additional paths to control the TEMs and TCMs in the real-space descendant system. Such capability is desirable for applications utilizing these modes.

In conclusion, we have demonstrated with both theory and acoustic experiments a 4D Chern and HOTI. Our work expands the concept of HOTIs to 4D systems. The ideas demonstrated in this paper are general and can be adapted for other types of waves, such as mechanical systems, electromagnetism, photonics, and cold atom systems. We can also expect rich phenomena to be discovered by the clever design of the modulation functions or by using other types of topologically nontrivial models. It can also be useful for building systems in even higher dimensions.

**Acknowledgments**

The authors thank Shiqiao Wu and Weiyuan Tang for sample preparation. Z.-G. C. and G. M. thank C. T. Chan, Zhao-Qing Zhang, Cheng He, Ruoyang Zhang, and Biao Yang for fruitful discussions. This work is supported by National Science Foundation of China (NSFC) Excellent Young Scientist Scheme (Hong Kong & Macau)  (#11922416), and NSFC Youth Program (#11802256), Hong Kong Research Grants Council (GRF 12300419, ECS 22302718, C6013-18G), and Hong Kong Baptist University (FRG2/17-18/056, RC-SGT2/18-19/SCI/006).

**References**

[1]   X.-L. Qi and S.-C. Zhang, Topological insulators and superconductors, Rev. Mod. Phys. **83**, 1057 (2011).
[2]   M. Z. Hasan and C. L. Kane, Colloquium: Topological insulators, Rev. Mod. Phys. **82**, 3045 (2010).
[3]   L. Lu, J. D. Joannopoulos, and M. Soljacic, Topological photonics, Nat. Photon, **8**, 821 (2014).
[4]   T. Ozawa *et al.*, Topological photonics, Rev. Mod. Phys. **91**, 015006 (2019).
[5]   N. R. Cooper, J. Dalibard, and I. B. Spielman, Topological bands for ultracold atoms, Rev. Mod. Phys. **91**, 015005 (2019).
[6]   G. Ma, M. Xiao, and C. T. Chan, Topological phases in acoustic and mechanical systems, Nat. Rev. Phys. **1**, 281 (2019).
[7]   X. Zhang, M. Xiao, Y. Cheng, M.-H. Lu, and J. Christensen, Topological sound, Commun. Phys. **1**, 97 (2018).
[8]   W. A. Benalcazar, B. A. Bernevig, and T. L. Hughes, Quantized electric multipole insulators, Science **357**, 61 (2017).
[9]   J. Langbehn, Y. Peng, L. Trifunovic, F. von Oppen, and P. W. Brouwer, Reflection-Symmetric Second-Order Topological Insulators and Superconductors, Phys. Rev. Lett. **119**, 246401 (2017).
[10]   Z. Song, Z. Fang, and C. Fang, d-2-Dimensional Edge States of Rotation Symmetry Protected Topological States, Phys. Rev. Lett. **119**, 246402 (2017).
[11]   M. Ezawa, Higher-Order Topological Insulators and Semimetals on the Breathing Kagome and Pyrochlore Lattices, Phys. Rev. Lett. **120**, 026801 (2018).
[12]   S. Imhof *et al.*, Topolectrical-circuit realization of topological corner modes, Nat. Phys. **14**, 925 (2018).
[13]   C. W. Peterson, W. A. Benalcazar, T. L. Hughes, and G. Bahl, A quantized microwave quadrupole insulator with topologically protected corner states, Nature **555**, 346 (2018).
[14]   F. Schindler, A. M. Cook, M. G. Vergniory, Z. Wang, S. S. P. Parkin, B. A. Bernevig, and T. Neupert, Higher-order topological insulators, Sci. Adv. **4**, 0346 (2018).
[15]   M. Serra-Garcia, V. Peri, R. Süsstrunk, O. R. Bilal, T. Larsen, L. G. Villanueva, and S. D. Huber, Observation of a phononic quadrupole topological insulator, Nature **555**, 342 (2018).
[16]   B.-Y. Xie, H.-F. Wang, H.-X. Wang, X.-Y. Zhu, J.-H. Jiang, M. H. Lu, and Y. F. Chen, Second-order photonic topological insulator with corner states, Phys. Rev. B **98**, 205147 (2018).
[17]   H. Xue, Y. Yang, F. Gao, Y. Chong, and B. Zhang, Acoustic higher-order topological insulator on a kagome lattice, Nat. Mater.  (2018).

[18] Z.-G. Chen, C. Xu, R. Al Jahdali, J. Mei, and Y. Wu, Corner states in a second-order acoustic topological insulator as bound states in the continuum, Phys. Rev. B **100**, 075120 (2019).
[19] X. Ni, M. Weiner, A. Alù, and A. B. Khanikaev, Observation of higher-order topological acoustic states protected by generalized chiral symmetry, Nat. Mater. **18**, 113 (2019).
[20] Z. Zhang, H. Long, C. Liu, C. Shao, Y. Cheng, X. Liu, and J. Christensen, Deep-Subwavelength Holey Acoustic Second-Order Topological Insulators, Adv. Mater., 1904682 (2019).
[21] W. Ma *et al.*, Experimental Observation of a Generalized Thouless Pump with a Single Spin, Phys. Rev. Lett. **120**, 120501 (2018).
[22] J. Noh, W. A. Benalcazar, S. Huang, M. J. Collins, K. P. Chen, T. L. Hughes, and M. C. Rechtsman, Topological protection of photonic mid-gap defect modes, Nat. Photon, **12**, 408 (2018).
[23] X. Zhang, H.-X. Wang, Z.-K. Lin, Y. Tian, B. Xie, M.-H. Lu, Y.-F. Chen, and J.-H. Jiang, Second-order topology and multidimensional topological transitions in sonic crystals, Nat. Phys. **15**, 582 (2019).
[24] L.-J. Lang, X. Cai, and S. Chen, Edge States and Topological Phases in One-Dimensional Optical Superlattices, Phys. Rev. Lett. **108**, 220401 (2012).
[25] Y. E. Kraus, Z. Ringel, and O. Zilberberg, Four-dimensional quantum Hall effect in a two-dimensional quasicrystal, Phys. Rev. Lett. **111**, 226401 (2013).
[26] Y. E. Kraus and O. Zilberberg, Topological Equivalence between the Fibonacci Quasicrystal and the Harper Model, Phys. Rev. Lett. **109**, 116404 (2012).
[27] W. Zhu, X. Fang, D. Li, Y. Sun, Y. Li, Y. Jing, and H. Chen, Simultaneous Observation of a Topological Edge State and Exceptional Point in an Open and Non-Hermitian Acoustic System, Phys. Rev. Lett. **121**, 124501 (2018).
[28] D. J. Apigo, W. Cheng, K. F. Dobiszewski, E. Prodan, and C. Prodan, Observation of Topological Edge Modes in a Quasiperiodic Acoustic Waveguide, Phys. Rev. Lett. **122**, 095501 (2019).
[29] Y. Long and J. Ren, Floquet topological acoustic resonators and acoustic Thouless pumping, J. Acoust. Soc. Am **146**, 742 (2019).
[30] L. Yuan, Q. Lin, M. Xiao, and S. Fan, Synthetic dimension in photonics, Optica **5**, 1396 (2018).
[31] X. Ni, K. Chen, M. Weiner, D. J. Apigo, C. Prodan, A. Alù, E. Prodan, and A. B. Khanikaev, Observation of Hofstadter butterfly and topological edge states in reconfigurable quasi-periodic acoustic crystals, Commun. Phys. **2**, 55 (2019).
[32] Q. Wang, M. Xiao, H. Liu, S. Zhu, and C. T. Chan, Optical Interface States Protected by Synthetic Weyl Points, Phys. Rev. X **7**, 031032 (2017).
[33] X. Fan, C. Qiu, Y. Shen, H. He, M. Xiao, M. Ke, and Z. Liu, Probing Weyl Physics with One-Dimensional Sonic Crystals, Phys. Rev. Lett. **122**, 136802 (2019).
[34] M. Lohse, C. Schweizer, H. M. Price, O. Zilberberg, and I. Bloch, Exploring 4D quantum Hall physics with a 2D topological charge pump, Nature **553**, 55 (2018).
[35] O. Zilberberg, S. Huang, J. Guglielmon, M. Wang, K. P. Chen, Y. E. Kraus, and M. C. Rechtsman, Photonic topological boundary pumping as a probe of 4D quantum Hall physics, Nature **553**, 59 (2018).
[36] S.-C. Zhang and J. Hu, A Four-Dimensional Generalization of the Quantum Hall Effect, Science **294**, 823 (2001).
[37] C. W. Hsu, B. Zhen, A. D. Stone, J. D. Joannopoulos, and M. Soljačić, Bound states in the continuum, Nat. Rev. Mater. **1**, 16048 (2016).
[38] See Supplemental Material for additional details on the correspondence between the tight-binding models and our acoustic structures; parameter discussion; the topological characterizations and robustness discussion of the THMs and TSMs; construct rules; the simulation results of the boundary modes; two experimental reults involving zero modulation phases.
[39] X.-L. Qi, T. L. Hughes, and S.-C. Zhang, Topological field theory of time-reversal invariant insulators, Phys. Rev. B **78**, 195424 (2008).
[40] M. Mochol-Grzelak, A. Dauphin, A. Celi, and M. Lewenstein, Efficient algorithm to compute the second Chern number in four dimensional systems, Quantum Science and Technology **4**, 014009 (2018).
[41] A. M. Marques and R. G. Dias, Analytical solution of open crystalline linear 1D tight-binding models, J. Phys. A: Math. Theor. **53**, 075303 (2020).
[42] X. Zhang, Y. Chen, Y. Wang, Y. Liu, J. Y. Lin, N. C. Hu, B. Guan, and C. H. Lee, Entangled four-dimensional multicomponent topological states from photonic crystal defects, Phys. Rev. B **100**, 041110 (2019).
[43] Y.-X. Xiao, G. Ma, Z.-Q. Zhang, and C. T. Chan, Topological Subspace-Induced Bound State in the Continuum, Phys. Rev. Lett. **118**, 166803 (2017).

[44]     N. Kaina, F. Lemoult, M. Fink, and G. Lerosey, Negative refractive index and acoustic superlens from multiple scattering in single negative metamaterials, Nature **525**, 77 (2015).
[45]     M. Robnik, A simple separable Hamiltonian having bound states in the continuum, J. Phys. A: Math. Gen. **19**, 3845 (1986).

# Supplemental Materials
# "Four-Dimensional Higher-Order Chern Insulator and Its Acoustic Realization"

Ze-Guo Chen[1], Weiwei Zhu[1,2], Yang Tan[1,3], Licheng Wang[1,4], and Guancong Ma[1,*]

[1]Department of Physics, Hong Kong Baptist University, Kowloon Tong, Hong Kong

**Contents**

# I. Acoustic realization and tight-binding model fitting

Coupled acoustic cavities are an excellent platform for realizing a tight-binding model [1-3]. Our acoustic system consists of cylindrical acoustic cavities connected by small coupling waveguides. The synthetic dimensions which take effect as onsite modulation can be easily realized by adjusting the height of the cavities. To show this, we analyze two cavities connected by one rectangular tube, as illustrated in Fig. S1(a). This system can be understood by a tight-binding model and characterized by a 2×2 Hamiltonian: $H = \begin{bmatrix} f(h) & t \\ t & f(h) \end{bmatrix}$. We further let $f(h) \sim 1/h$, since the onsite eigenmode is the fundamental cavity mode with the resonant frequency proportional to $1/h$. The calculated eigenfrequency is compared with the numerical simulation to obtain fitting parameters $t = -124.75$ Hz, $f(h) = (126.46 + 168.36/h)$ Hz. The negative sign comes from the fact that symmetric mode has a lower frequency than anti-symmetric modes. Excellent agreement between tight-binding theory and simulation is obtained (Fig. S1(b)), which validates our acoustic approach.

The 4D system is realized by choosing $\phi_x, \phi_y$. In a 2D acoustic lattice, this is done by tuning the height of each cavity. The onsite eigenfrequencies of the 2D acoustic lattice are given by

$$f_{m,n}(\phi_x, \phi_y) = f_0 + \lambda_x \cos(2\pi b_x m + \phi_x) + \lambda_y \cos(2\pi b_y n + \phi_y)$$

The height $h_{m,n}$ of the cavity labeled $(m,n)$ is tuned according to $f_{m,n}(h_{m,n}) = (126.46 + 168.36/h_{m,n})$ Hz. The resulted heights of the cylindrical cavities fall in the region between 71.6 mm and 120mm.

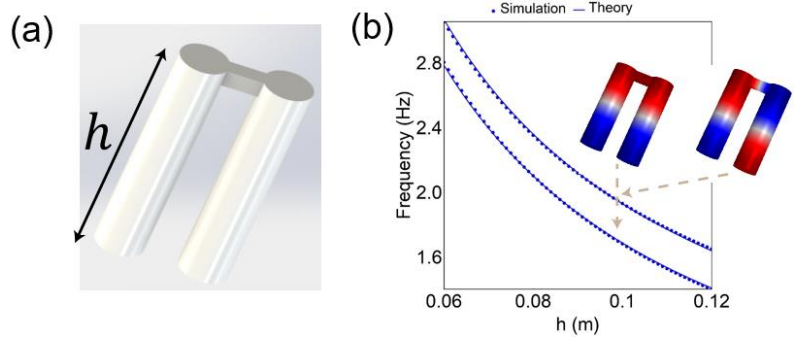

**Figure S1** (a) The schematic drawing of two coupled acoustic cavities with tunable height $h$. (b) The simulated (blue dots) and theoretic (red dots) eigenfrequencies of the double cavities. The inset shows the two modes.

It should be noted that our acoustic lattice has eigenspectra that slightly deviate from an ideal tight-binding model. Our investigation shows that such deviation is mainly due to an additional perturbation introduced by the coupling waveguides to the onsite resonant frequency of the cavities. This perturbation is linearly proportional to the number of waveguides connected to the cavity. It can be seen for the 1D chain, each bulk cavity has two coupling waveguides, and the cavities at the boundaries are connected to only one. We found that the perturbations to bulk/boundary cavities are 95 Hz and 190 Hz, respectively. Accounting for these perturbations in the tight-binding model, band structures that excellently match to the acoustic results can be obtained, as shown in Fig. S2(b) and (c). Likewise, for the 2D lattice, the corner/edge/bulk sites which are respectively connected to two/three/four coupling waveguides, the perturbations are 95/190/285 Hz. Excellent agreement with acoustic eigenspectra can be obtained using the modified tight-binding model, which has taken the onsite perturbations into account (Fig. 3 in the main text).

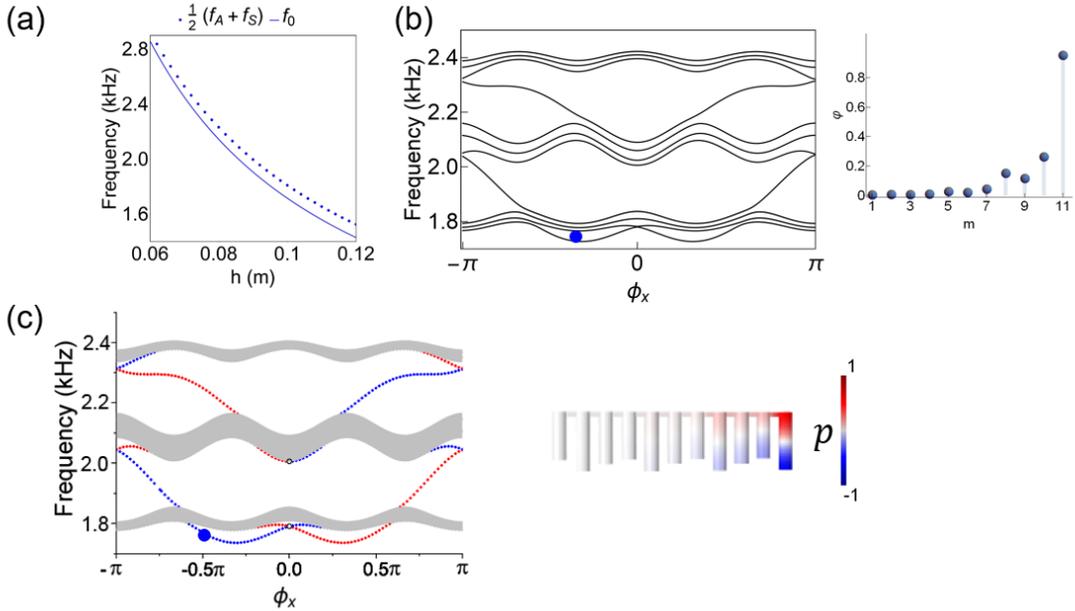

**Figure S2** (a) The onsite frequency in tight-binding models compared with the first-order resonant frequency of an isolated cavity. (b) The eigenfrequency spectra in a modified 1D tight-binding model. The right inset shows boundary mode as an extension of gapless mode in the first gap. (c) The eigenfrequency spectra computed by COMSOL. Compared to the results in (b), excellent agreement is seen.

## II. System parameters and their effects

Here, we discuss several important parameters in our system. To keep the discussion concise, we use the 1D chain

$$\hat{H} = \sum_m^N [f_m(\phi_x)|m\rangle\langle m| + t|m\rangle\langle m+1| + t|m+1\rangle\langle m|] \quad (1)$$

with $f_m(\phi_x) = f_0 + \lambda_x \cos(2\pi m b_x + \phi_x)$. We focus on the effect of three parameters: the modulation frequency $b_x$, the total site number $N$, and the modulation amplitude $\lambda_x$.

The modulation frequency $b_x = p/q$ with $p$ and $q$ being co-prime is a rational number in our model. This means that the system is periodic, and each unit cell contains $q$ sites. As a result, the system band structure has $q$ bulk bands, as shown in Fig. S3(a) for $b_x = \frac{1}{3}, \frac{1}{4}, \frac{2}{5}$. For a finite system with $N$ sites, the $b$ can be continuously tuned. We set $N = 32$, and the eigenfrequency spectra as a function of $b_x$ show a typical Hofstadter butterfly structure [4] (Fig. S3(b)).

The number of sites $N$ affects the position of the topological boundary modes in the synthetic dimension $\phi_x$. This is shown in Fig. S3(c). Note that the existence and the number of the boundary modes are protected by nonzero Chern numbers and are not affected by the site number. In general, a rational modulation frequency $b_x = p/q$, the band structure, is symmetric about $\phi_x = 0$ when the lattice size is $N = nq - 1$, where $n$ is an integer. This can be theoretically shown by applying a unitary operator $P$, defined as $P_{ij} = 1$ for $i + j = N + 1$, which induces $\hat{H}(\phi_x) = P^{-1}\hat{H}(-\phi_x)P$. In our realization, we choose 11 cavities which belong to the case of $N = nq - 1$, wherein the boundary modes are symmetric about $\phi_x = 0$ so that they do not cross. This characteristic is optimal for experimental observation of the TCMs and TEMs, since the crossing of the topological modes can be avoided.

The modulation amplitude $\lambda_x$ mainly affects the size of the bandgaps, as shown in Fig. S3(d). A small $\lambda_x$ results in small bandgaps, which will make the in-gap phenomena hard to observe, especially in the presence of loss. In addition, small modulation to the onsite requires higher accuracy in adjusting the cavity heights, which increases the experimental difficulty. On the other hand, a large $\lambda_x$ means that the cavity height must be varied to a larger degree, which decreases the fidelity of the acoustic lattice in reproducing the tight-binding model.

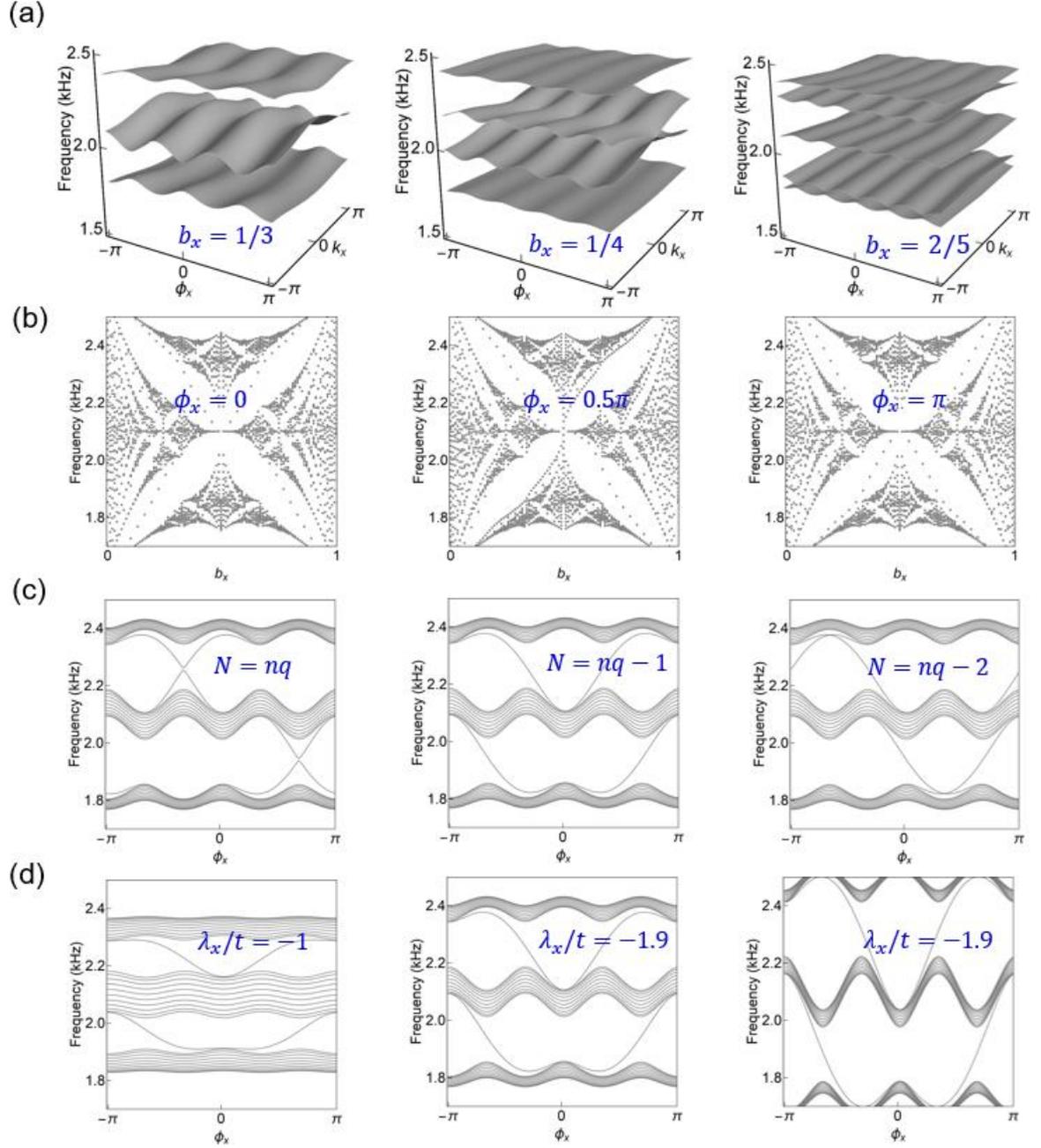

**Figure S3** (a) The bulk band structure with different values of modulation frequency $b_x = 1/3,\ 1/4,\ 2/5$, where $\lambda_x/t = -1.9$. (b) The eigenspectra as functions of $b_x$ for a finite system with $N = 32$, $\lambda_x/t = -1.9$. (c) The eigenspectra's dependence of the site number $N$. (d) The effect of modulation amplitude $\lambda_x$.

### III.  2D Chern insulator from a 1D chain

The Hamiltonian of the 1D AAH model is given by Eq. (1), wherein $f_m(\phi_x) = f_0 + \lambda_x \cos(2\pi m b_x + \phi_x)$ is the onsite frequency modulation, with $b_x$ and $\phi_x$ respectively being the modulation frequency and modulation phase factor. For any given $\phi_x$, Eq. (1) can be viewed as a Fourier transformation of another specific 2D Hamiltonian in $\phi_x$. To show this, we rewrite the notation $|m\rangle$ in 1D Hamiltonian as $|m,\phi_x\rangle$, which obeys the commutation relation $\{\langle m',\phi'_x|,|m,\phi_x\rangle\} = \delta_{m,m'}\delta_{\phi_x,\phi'_x}$. The Fourier transformation of $|m,\phi_x\rangle$ leads to $|m,\phi_x\rangle = \sum_n e^{-i\phi_x l}|m,l\rangle$, which means the phase factor $\phi$ is viewed as the pseudo-momentum of an additional dimension with a lattice labeled by $l$. Then we can write the 2D Hamiltonian as $\mathcal{H} = \int_0^{2\pi}(d\phi_x/2\pi)\widehat{H}(\phi_x)$, the expression is [5]:

$$\mathcal{H}(m,l) = \sum_{m,l} f_0|m,l\rangle\langle m,l| + \sum_{m,l}\left(t|m,l\rangle\langle m+1,l| + \frac{\lambda_x e^{i2\pi b_x m}}{2}|m,l\rangle\langle m,l+1| + \text{H.c.}\right). \quad (2)$$

It is easy to see that $\mathcal{H}(m,l)$ describe a 2D rectangular lattice in which each unit cell exhibits a uniform magnetic flux $b_x$, where the magnetic field appears in a Landau gauge.

Note that the modulation frequency $b_x = 1/q$ (with $q$ being an integer) indicates that each unit cell contains $q$ sites. Consequently, the system's eigenspectrum has $q$ bands. In Fig. S4(a), we show the calculated band structure of an infinite three-site lattice with $b_x = 1/3$. The Chern number of each band can be calculated in the synthetic plane of $(k_x, \phi_x)$

$$C_B = \frac{i}{2\pi}\int dk_x d\phi_x \, \text{Tr}\left[P\left(\frac{\partial P}{\partial k_x}, \frac{\partial P}{\partial \phi_x}\right)\right], \quad (3)$$

where $P$ is the projected matrix of the occupied bands, Tr denotes the trace of the matrix. The results for the three bands are $(1,-2,1)$, respectively.

We now consider a finite 1D chain for $b_x = 1/3$ with $N$ sites. The system has a total of $N$ eigenmodes, as illustrated in Fig. S4(a). However, most of these $N$ modes are clustered into three groups, separated by two well-defined bandgaps. These are identified to be bulk modes. On the other hand, the modes that connect the bandgaps are boundary modes. Therefore, it is informative to compute the gap Chern number $C_G$, which is the sum of the band Chern numbers below the gap. For $b_x = 1/3$, the gap Chern numbers are $(1,-1)$.

We further verify our calculated gap Chern numbers using an analytical formula. For a rational flux $b_x = p/q$, the gap of the 1D AAH model is characterized by two integers $s_r$ and $C_{G,r}$, which are determined by a Diophantine equation [6]:

$$r = qs_r + pC_{G,r} \quad |C_{G,r}| \leq q/2, \quad s_r, C_{G,r} \in \mathbb{Z}, \tag{4}$$

where $r$ denotes the $r$th gap. The equation's solutions, $s_r, C_{G,r}$, are uniquely defined. $C_{G,r}$ is the first Chern number for the $r$th bandgap. For our three-band 1D chain ($b_x = 1/3$ with $q = 3, p = 1$), the solutions for the first bandgap ($r = 1$) are $(s_{r=1}, C_{G,r=1}) = (0, 1)$, so that Eq. (4) is satisfied as $1 = 3 \times 0 + 1 \times 1$. For the second bandgap, $r = 2$, we have $2 = 3s_{r=2} + C_{G,r=2}$, so that $(s_{r=2}, C_{G,r=2}) = (1, -1)$. Therefore, the Chern numbers of the two bandgaps are $C_{G,r=1} = 1$ and $C_{G,r=2} = -1$, respectively, which are consistent with the numerical results shown above.

The similar calculations are done for a six-site lattice with $b_x = 1/6$. As expected, six bands are found. However, since the third and fourth bands touch at certain values of $(k_x, \phi_x)$, we can obtain five Chern numbers: $(1, 1, -4, 1, 1)$. The third and fourth bands have a total Chern number of $-4$. These are shown in Fig. S4(b). The gap Chern numbers are consequently $(1, 2, -2, -1)$ for the four complete bandgaps.

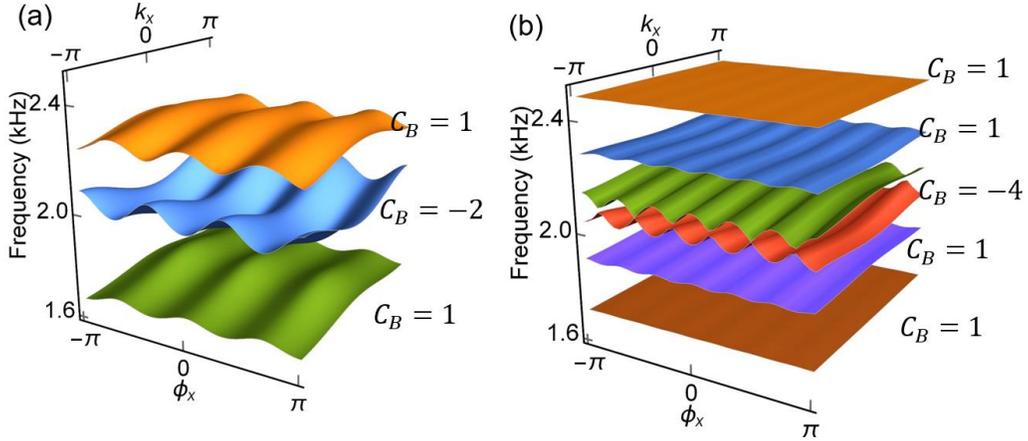

**Figure S4** The band structures for an infinite 2D Chern system as functions of Bloch wavevector $k_x$ and the modulation phase $\phi_x$. (a) The band structure of the 1D system with $b_x = 1/3$. (b) The band structure of the 1D system with $b_x = 1/6$. We mark the first Chern number for each isolated band $C_B$. In (b), the third and fourth bands touch at $k_x = \pm\pi$ for several values of $\phi_x$. These two bands together have a first Chern number $C_B = -4$.

### IV. Topological invariant of the TSMs

Our analysis shows that the TCMs are the real-space descendants of (*4-2*)-dimensional second-order topological surface modes (TSMs) of the 4D system. As shown in Fig. 2(b-d) in the main text, these TSMs close the bandgaps of (*4-1*)-dimensional first-order THMs, indicating that they are topologically non-trivial. The TSMs are protected by the nonzero combinations of two first Chern numbers $\mathcal{C} \equiv (C_G^x, C_G^y)$, where $C_G^x$ and $C_G^y$ are defined for 2D sub-systems existing on $k_x\phi_x$ and $k_y\phi_y$-planes, respectively. Here, we further show that $\mathcal{C} \equiv (C_G^x, C_G^y)$ is linked to the second Chern number of the 4D system.

Simply put, the second Chern number of the 4D bandgap with energy around $\epsilon$ is related to the first Chern numbers of bands of 2D sub-systems with energy $\epsilon_x + \epsilon_y < \epsilon$.

$$\mathbb{C}_{G,\epsilon} = \sum_{\epsilon_x+\epsilon_y<\epsilon} C^x_{B,\epsilon_x} C^y_{B,\epsilon_y}. \tag{5}$$

Eq. (5) is a proven relation [7] between the first Chern numbers for bands and the second Chern number for gaps. It can be extended by relating the first band Chern numbers $C_B^x$, $C_B^y$ to the first gap Chern numbers $C_G^x$, $C_G^y$. To explicitly show this, we mark the second Chern number for the $r$th 4D bandgap as $\mathbb{C}_{G,r}$, then we can write

$$\mathbb{C}_{G,\epsilon_{r+1}} = \mathbb{C}_{G,\epsilon_r} + \sum_{\epsilon_r<\epsilon_x+\epsilon_y<\epsilon_{r+1}} C^x_{B,\epsilon_x} C^y_{B,\epsilon_y}, \tag{6}$$

wherein $\epsilon_r$ is an energy value within the $m$th 4D bandgap. Summation of the left-hand-side from 1 to $m$, and we have

$$\sum_{i=1}^r \mathbb{C}_{G,\epsilon_{i+1}} = \sum_{i=1}^r \mathbb{C}_{G,\epsilon_i} + \sum_{\epsilon_x+\epsilon_y<\epsilon_{r+1}} C^x_{B,\epsilon_x} C^y_{B,\epsilon_y}. \tag{7}$$

Next, we use mathematical induction to prove: $\sum_{i=1}^r \mathbb{C}_{G,\epsilon_i} = \sum_{\epsilon_{r-1}<\epsilon_x+\epsilon_y<\epsilon_r} C^x_{G,\epsilon_x} C^y_{G,\epsilon_y}$, where $C^x_{G,\epsilon_x}$ is the first Chern number for the bandgap with the energy not larger than $\epsilon_x$. This relation obviously holds for $r = 1$, since for the 1st band and bandgap of a 2D Chern insulator, there must be $C_{G,1} = C_{B,1}$, and therefore

$$\mathbb{C}_{G,\epsilon_1} = \sum_{\epsilon_0<\epsilon_x+\epsilon_y<\epsilon_1} C^x_{B,\epsilon_x} C^y_{B,\epsilon_y} = \sum_{\epsilon_0<\epsilon_x+\epsilon_y<\epsilon_1} C^{\epsilon_x}_G C^{\epsilon_y}_G. \tag{7}$$

Next, if the rule holds for arbitrary $r$, then it holds for $r + 1$, viz.

$$\sum_{i=1}^r \mathbb{C}_{G,\epsilon_{i+1}} = \sum_{\epsilon_{r-1}<\epsilon_x+\epsilon_y<\epsilon_r} C^x_{G,\epsilon_x} C^y_{G,\epsilon_y} + \sum_{\epsilon_x+\epsilon_y<\epsilon_{r+1}} C^x_{B,\epsilon_x} C^y_{B,\epsilon_y}$$

$$= \sum_{\epsilon_{r-1}<\epsilon_x+\epsilon_y<\epsilon_r} C^x_{G,\epsilon_x} C^y_{G,\epsilon_y} + \sum_{\epsilon_r<\epsilon_x+\epsilon_y<\epsilon_{r+1}} C^x_{B,\epsilon_x} C^y_{G,\epsilon_y}$$

$$= C^{\epsilon_y}_G \left( \sum_{\epsilon_{r-1}<\epsilon_x+\epsilon_y<\epsilon_{r+1}} C^x_{G,\epsilon_x} + \sum_{\epsilon_r<\epsilon_x+\epsilon_y<\epsilon_{r+1}} C^x_{B,\epsilon_x} \right)$$

$$= \sum_{\epsilon_r<\epsilon_x+\epsilon_y<\epsilon_{r+1}} C^x_{G,\epsilon_x} C^y_{G,\epsilon_y}.$$

The second-last step utilizes the relation that the gap Chern number is the sum of all band Chern numbers below the bandgap. $C_G^{i+1} = C_G^i + C_B^{i+1}$.

The above shows that the second Chern numbers for 4D bandgaps are tied to the first gap Chern numbers of the 2D Chern insulator sub-systems:

$$\sum_{i=1}^{r} \mathbb{C}_{G,\epsilon_i} = \sum_{\epsilon_{r-1}<\epsilon_x+\epsilon_y<\epsilon_r} C_{G,\epsilon_x}^x C_{G,\epsilon_y}^y. \tag{8}$$

Eq. (8) means that the topological invariant $\mathcal{C} \equiv (C_G^x, C_G^y)$ is indeed related to the second Chern number. Consequently, the (*4-2*)-dimensional TSMs, and thereby the TCMs, are unquestionably protected by 4D topology.

## V.     Robustness against disorders

Due to the topological protection, the 4D system and the boundary modes therein are robust against disorder. To show this, we introduce disorders to the system in the form of random onsite perturbations $\delta f = s\lambda \, \text{rand}(-1, 1)$, wherein $s$ is the perturbation strength, and $\lambda = \lambda_x + \lambda_y$ is the sum of modulation amplitudes in both spatial directions. We observe that THMs and TSMs are robust against relatively large disorders. An example is shown in Fig. S5. In Fig. S5(a), wherein $s = 0.1$, the gapless THMs and TSMs are clearly identified. Fig. S5(b) shows that the THMs are well localized on the edges, despite the disorders. Likewise, Fig. S5(c) shows that the TSMs localize at the corners. Hence these results are also further evidence of the topologically nontrivial characteristics of the TCMs and TEMs.

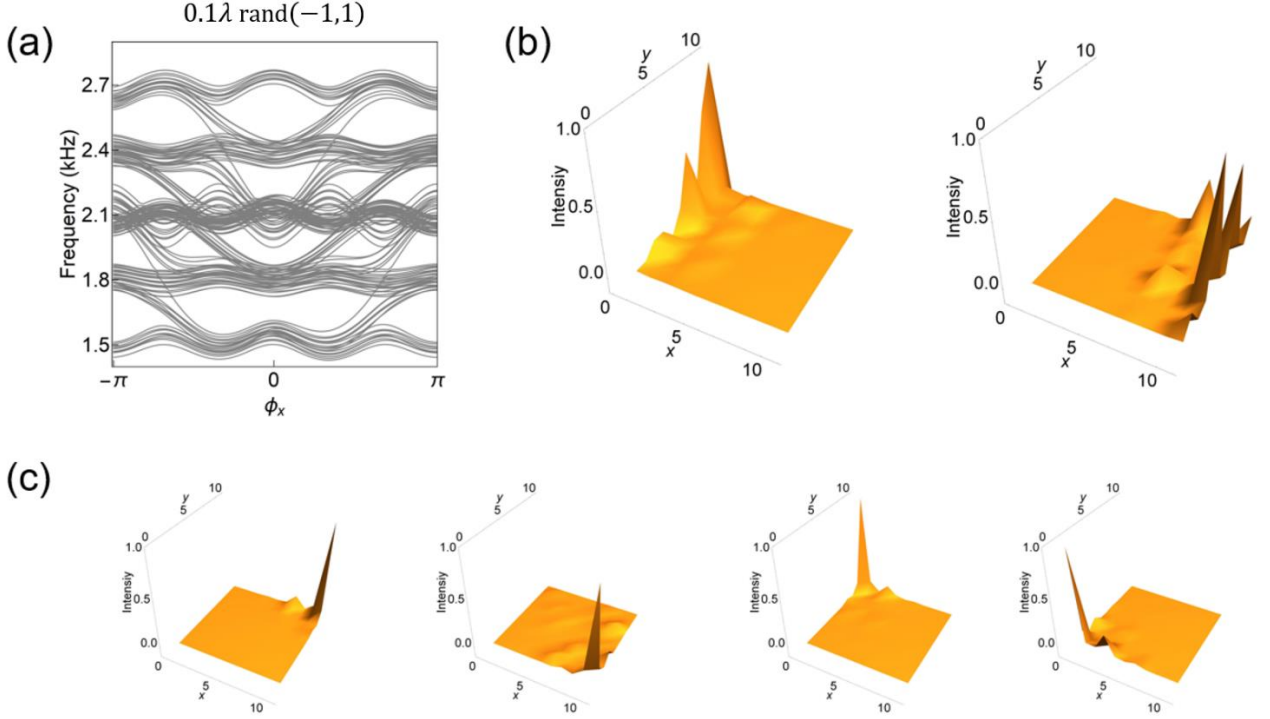

**Figure S5** (a) The eigenfrequency spectra under perturbations $\delta f = 0.1\lambda \text{ rand}(-1,1)$. The results show gapless THMs and TSMs. Real-space field distributions of THMs indicate that they are well localized at the edges (b), and the TSMs are strongly confined at the corners (c).

## VI. Topological boundary modes with a modulation frequency $b_x = 1/6$

In this section, we analyze the boundary modes of a 2D Chern insulator based on a finite 1D AAH chain with a modulation frequency $b_x = 1/6$. It is easy to see that there are six bulk bands instead of three, which is clearly seen from the eigenspectra obtained from the tight-binding calculation of a 65-site chain (Fig. S6(a)). As shown in section II, the system is also a Chern insulator, with gap Chern numbers labeled in Fig. S6(a). We mark the synthetic coordinate $\phi_x = -0.6\pi$, at which the system has five boundary modes, labeled by the letters a – e. Modes-a, b, c are localized at the left boundary. Modes-d, e are localized at the right boundary. Here, mode-a and mode-c deserve more discussion. For mode-a, the largest sound amplitude is found not at the boundary cavity, but at the cavity one site away. Such a distribution directly affects one of the profiles of the corresponding TCM, which is also the reason why we need to place the source one site away from the corner for the best result at the frequency $f_{II}^3$, as indicated in Fig. 6(c) in the main text. On the other hand, mode-c is found in the third bandgap, which is an incomplete bandgap in the $k_x\phi_x$-plane. Our analysis shows that the degeneracy of the two touching bands (the third and the fourth bands) can be easily lifted by introducing small perturbations to onsite eigenfrequencies. The same also occurs when the lattice is finite. In fact, careful examination of the band structure of the acoustic model reveals the existence of a mini gap, which is attributed to the finite-sized effect of the lattice and the acoustic model's small deviation from an ideal tight-biding model. Nevertheless, we find neither the small perturbations nor the finite-sized effect has any pronounced effects on the existence of the branch of boundary modes that includes mode-$c$. In fact, the existence of the boundary mode branch is attributed to the nonzero Berry curvature distribution of the two sandwiching bands. This is, in a sense, similar to edge modes in valley Hall systems, which is underlain by nonzero local Berry curvature existing in the vicinity of a broken Dirac cone.

Eigenspectra obtained by finite-element simulations are shown in Fig. S6(b). The profiles of the boundary modes are plotted in the right panels. Excellent agreement with the tight-binding results is seen.

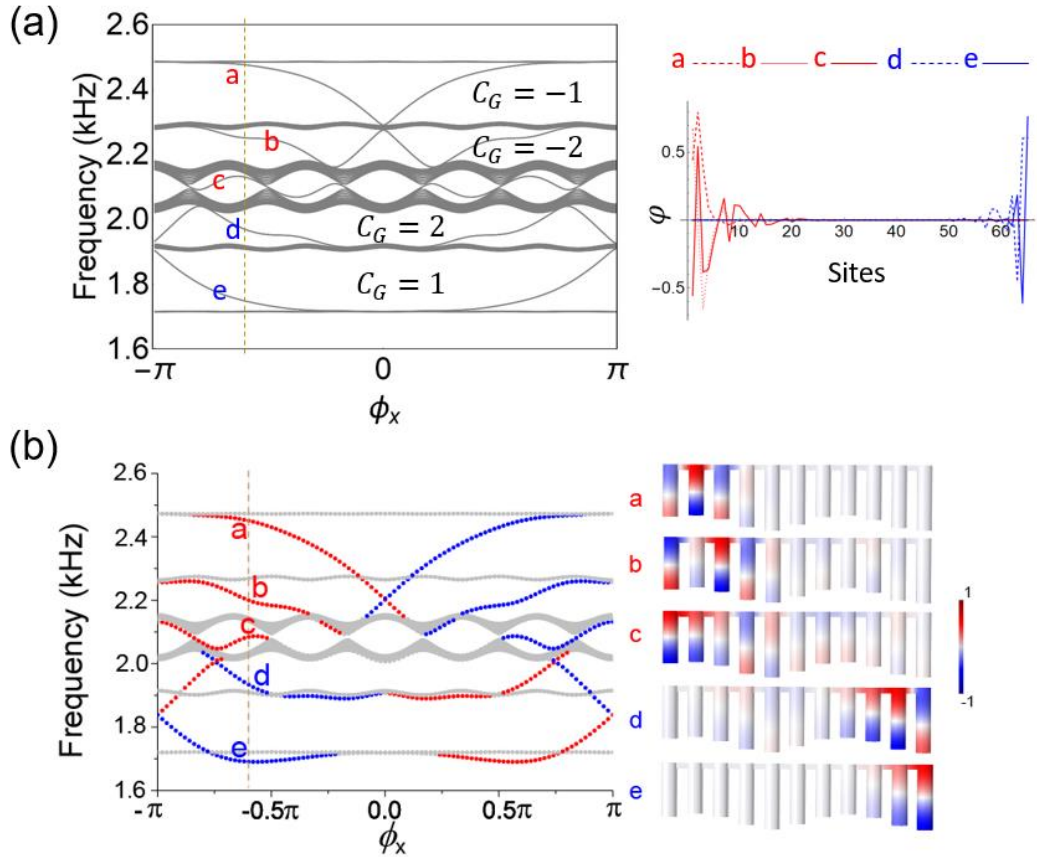

**Figure S6** (a) The calculated eigenspectra of a 1D AAH system with $b_x = 1/6$ using the tight-binding model. 65 sites are used in the calculation. Five boundary modes are found at $\phi_x = -0.6\pi$, which are labeled by the letters "a-e." Their eigenfields are also shown in the right panel. (b) The eigenspectra of the 1D acoustic chain as functions of $\phi_x$. The five boundary modes at $\phi_x = -0.6\pi$ are also marked and their eigenfields are shown in the right panel.

## VII. Formation rules of the eigenmodes of the 4D system

The 4D Hamiltonian $\mathbb{H}(\phi_x, \phi_y)$ can be viewed as two compounded copies of 2D Chern insulators using the mathematical operation of Kronecker product. It follows that the eigenmodes of $\mathbb{H}(\phi_x, \phi_y)$ can also be constructed as $|\Psi_{4D}\rangle = |\psi_y\rangle \otimes |\psi_x\rangle$, with $|\psi_x\rangle$ and $|\psi_y\rangle$ respectively being the eigenmodes of $H_x(\phi_x)$ and $H_y(\phi_y)$. By considering the characteristics of the 2D Chern model's eigenmodes, we arrive at three eigenmode formation rules for the 4D system: 4D bulk modes are the product of bulk modes in both $H_x(\phi_x)$ and $H_y(\phi_y)$, i. e., $|\Psi_{4D}^{bulk}\rangle = |\psi_y^{bulk}\rangle \otimes |\psi_x^{bulk}\rangle$; THMs, if any, are the product of 2D bulk modes and boundary modes of $|\Psi_{4D}^{THM}\rangle = |\psi_y^{boundary}\rangle \otimes |\psi_x^{bulk}\rangle$ or $|\Psi_{4D}^{THM}\rangle = |\psi_y^{bulk}\rangle \otimes |\psi_x^{boundary}\rangle$; TSMs, if exist, are composed of the product of two boundary modes $|\Psi_{4D}^{TSM}\rangle = |\psi_y^{boundary}\rangle \otimes |\psi_x^{boundary}\rangle$. This also indicates that the existence of the TSMs are the consequence of topological boundary modes in both the *x*- and *y*-directions 2D Chern insulator. This results also conforms well with the fact that TSMs are topologically protected by two nonzero first Chern numbers, since they each guarantees the existence of topological boundary modes in the Chern insulator in the 2D sub-dimensions. Meanwhile, these operations suggest that the eigenfrequencies of the 4D modes are given by the Minkowski sum of the 2D modes. Therefore the bulk modes' eigenfrequencies are $E_{4D}^{bulk} = E_x^{bulk} + E_y^{bulk} - f_0$, the THMs follow $E_{4D}^{THM} = E_x^{boundary} + E_y^{bulk} - f_0$ or $E_{2D}^{THM} = E_y^{boundary} + E_x^{bulk} - f_0$, and the TSMs follow $E_{4D}^{TSM} = E_x^{boundary} + E_y^{boundary} - f_0$.

To illustrate how these formation rules work, let us focus on the parametric point of $\phi_x = \phi_y = -0.5\pi$. We reproduce the 4D eigenspectra in Fig. S7(b). And the eigenspectra of the 2D constituent system is shown in Fig. S7(a). We recall that, at $\phi_x = -0.5\pi$, $H_x(\phi_x)$ and $H_y(\phi_y)$ possess three bulk bands and also topological boundary modes. It follows from eigenmode formation rules that the 4D system must have five bulk band regions, six THMs bands, and four TSMs. According to the distribution of boundary modes in the 2D model (Fig. S7(a)), we can precisely predict the location and frequency of THMs and TSMs.

The locations of the TSMs and in-gap THMs are indicated in the inset of Fig. S7(b). The THMs localized at edges parall to $x$ axis in real space are attributed to $|\Psi_{4D}^A\rangle = |\psi_y^\alpha\rangle \otimes |\psi_x^{bulk}\rangle$ and $|\Psi_{4D}^C\rangle = |\psi_y^\gamma\rangle \otimes |\psi_x^{bulk}\rangle$, wherein the superscripts $A, C$ label the edges in the 2D lattice, as shown in Fig. S7(b) and $|\psi_y^\alpha\rangle, |\psi_y^\gamma\rangle$ are respectively the boundary modes of the constituent 2D Chern insulator described by $H_y(\phi_y)$, which are shown in Fig. S7(a). Likewise, the THMs localized at edges parall to $y$ axis are attributed to $|\Psi_{4D}^B\rangle = |\psi_y^{bulk}\rangle \otimes |\psi_x^\alpha\rangle$ and $|\Psi_{2D}^D\rangle = |\psi_y^{bulk}\rangle \otimes |\psi_x^\gamma\rangle$. The four TSMs are: $|\Psi_{4D}^{III}\rangle = |\psi_y^\gamma\rangle \otimes |\psi_x^\gamma\rangle$ at the lowest frequency and $|\Psi_{4D}^I\rangle = |\psi_y^\alpha\rangle \otimes |\psi_x^\alpha\rangle$ at the highest frequency, sandwiching two degenerate modes $|\Psi_{4D}^{II}\rangle = |\psi_y^\gamma\rangle \otimes |\psi_x^\alpha\rangle$ and $|\Psi_{4D}^{IV}\rangle = |\psi_y^\alpha\rangle \otimes |\psi_x^\gamma\rangle$, where I, II, III, IV label the four corners as shown in Fig. S7(b). All the results are consistent with the construction rules.

The same eigenmode formation rules also suit the cases where $b_x \neq b_y$. We have shown in section VI that with $b_x = 1/6, \phi_x = -0.6\pi$, there are multiple boundary modes in one boundary, which accounts for multiple TSMs in the main text.

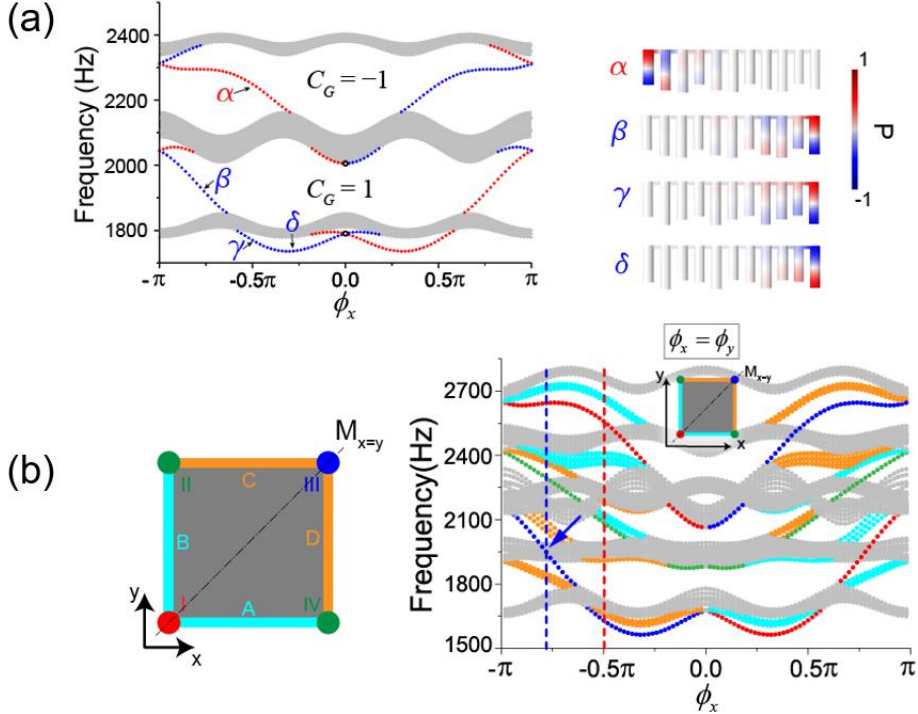

**Figure S7** Eigenmode formation rules in the acoustic lattice. (a) The eigenspectra of the acoustic Chern insulator from a 1D acoustic chain with $b_x = 1/3$ as a function of $\phi_x$. The right panels show the real-space distributions of several topological boundary modes. (b) The formations rules of the 4D eigenmodes enable us to determine the real-space location and frequencies of the bulk modes, THMs, and TSMs from the constituent 2D Chern insulators. The right panel shows the 4D eigenspectra along

the line of $\phi_x = \phi_y$, in which all modes are colored corresponding to their real-space locations, as shown in the left panel.

## VIII. Finite-element simulation of the 2D acoustic systems

First, we show the simulated results of the parametric point $(\phi_x, \phi_y) = (-0.5\pi, -0.5\pi)$ in Fig. S8. The four TCMs are clearly seen. These results and our experimental results (Fig. 3, main text) conform well.

Next, we discuss the simulated results of the 2D system with $b_x = 1/6, b_y = 1/3$. We set the system at $(\phi_x, \phi_y) = (-0.6\pi, -0.28\pi)$, which is the same as the experimental configuration. The calculated eigenspectrum is shown in Fig. 5(b). We re-plot it in Fig. S9, wherein we can identify three TCMs at corner II and two TCMs at corner III. This is consistent with our theoretical predictions and also aligns well with the experimental observations. We further obtain the field distribution patterns of these five TCMs. Noticeably, mode-$e$ has the maximum response amplitude one site away from the corner. This is consistent with our analysis in section IV. It is also the reason why we chose to excite mode-$e$ with a source located at the corresponding site, as shown in Fig. 5 (d) in the main text.

In Fig. S10, we show the simulated eigenspectra $b_x = b_y = 1/3$ at $\phi_y = 0$. The TEMs are highlighted. We can observe three groups of TEMs for both the left and the right edges. Note that the mode groups marked by "c" and "d" are TEMs as bound states in a continuum.

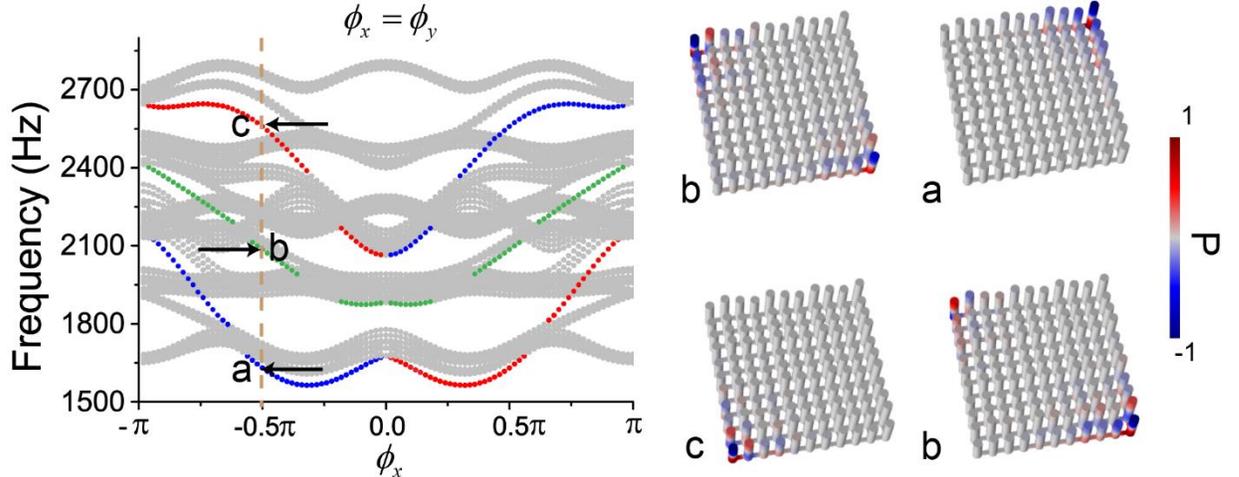

**Figure S8** The left panel shows the calculated eigenspectra of the 4D acoustic system with $b_x = b_y = 1/3$ as functions of $\phi_x$ along the parametric line of $\phi_x = \phi_y$. The simulated TSMs' eigenfields at $(\phi_x, \phi_y) = (-0.5\pi, -0.5\pi)$ at the frequencies marked by "a, b, c" are respectively shown in the right panels.

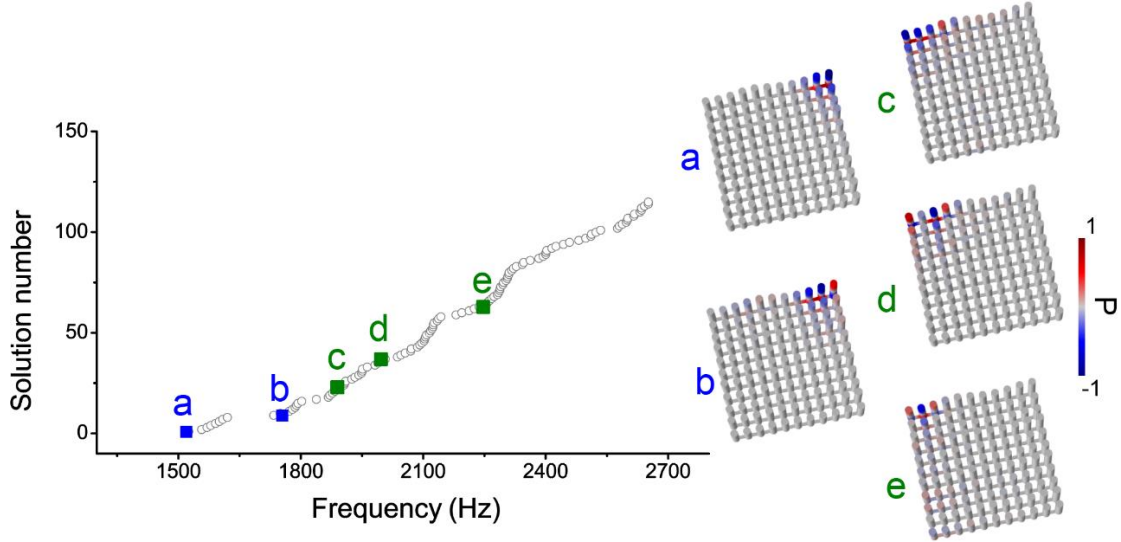

**Figure S9** The left panel shows the calculated eigenspectrum of the acoustic 4D system with $b_x = 1/6, b_y = 1/3$ at $(\phi_x, \phi_y) = (-0.6\pi, -0.28\pi)$, which is the same as the experimental configuration in the main text. Five TSMs are found. The modes in blue localize at corner III. The modes in green are at corner II. Their eigenfields are shown in the right panels.

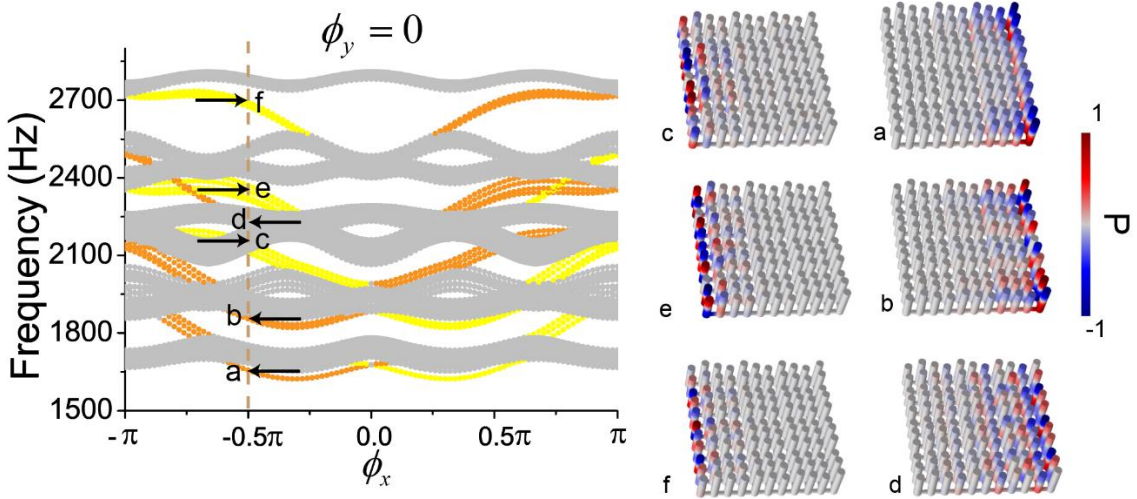

**Figure S10** The calculated eigenspectra of the 2D acoustic system with $\phi_y = 0$ as functions of $\phi_x$. For $\phi_x = -0.5\pi$, we show the simulated eigenfields of the THMs at frequencies marked by "a, b, c, d, e, f."

## IX. Experimental results at $(\phi_x, \phi_y) = (-0.5\pi, 0)$

Here, we show the experimental results of the 2D acoustic system at $(\phi_x, \phi_y) = (-0.5\pi, 0)$. At $\phi_x = 0$, the 1D chain does not possess well-distinguished boundary mode. From our formation rules (section VII), we know that: first, the 4D system only has THMs at edges B, D, but edges A and C do not sustain well-distinguished THM; second, TSM shall be absent.

Our predictions are experimentally verified. Note that the system has mirror symmetry

$M_x$. The response spectra measured in the bulk and at corners I and III (as defined in Fig. S11(a)) are shown in Fig. S11(b). The corner responses each have two resonant peaks. Note that response peaks can be due to either bulk modes, THMs at either joining edges (or both), or TSMs. To confirm the nature of these peaks, we raster-map the field distributions with a source at corner I and III. Both cases clearly show that the modes are confined along edges B and D, respectively (Fig. S11(c, d)), whereas no modes along edges A & C are seen. The symmetry $M_x$ suggests the same to be observed for corners II and IV. These observations align well with our prediction.

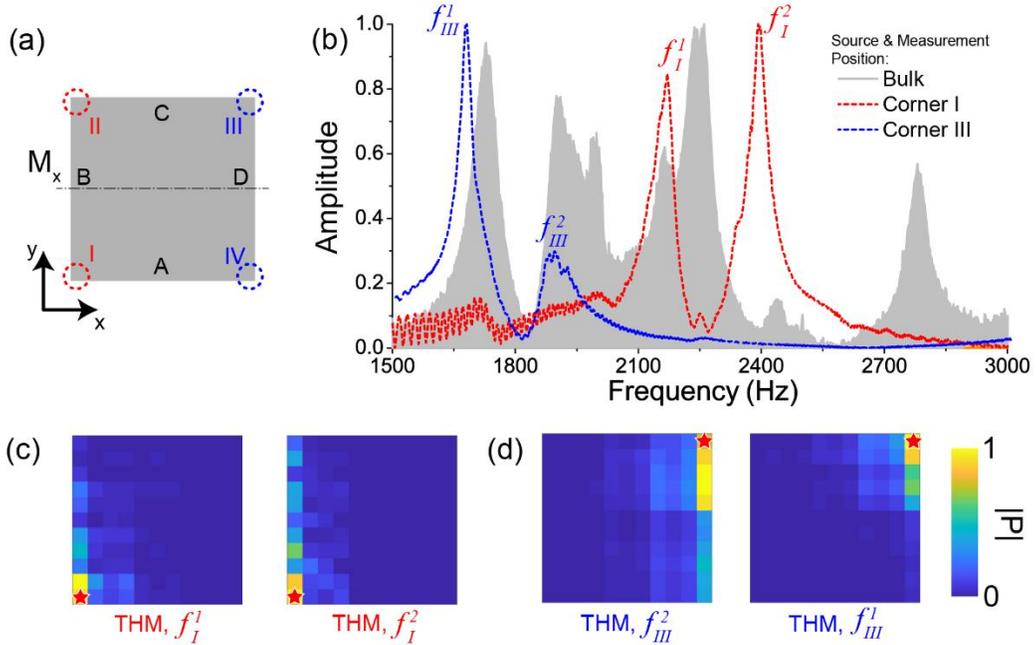

**Figure S11** The experimental results at $(\phi_x, \phi_y) = (-0.5\pi, 0)$. (a) A schematic drawing of the system, with the bulk, edges, corners marked in different colors. (b) The gray-colored region is pressure response in the bulk, the red/blue dashed curves are respectively the response measured at corner I and corner III. (c) The field maps when excited by a source located at corner I at the two frequencies as marked in (b). The distributions suggest THMs only found at edge B. (d) The field maps at two frequencies from the peak of the corner III response spectrum. The distributions suggest THMs localized at edge D. The red stars mark the source positions.

## X. Experimental results at $(\phi_x, \phi_y) = (0, 0)$

We tune the height of each cavity (by changing the amount of water) to drive the system to $(\phi_x, \phi_y) = (0, 0)$. From Fig. S7, we know that both THMs and TSMs shall be missing in the 4D system. Fig. S12(b) shows the measured response spectra, wherein the data are colored according to Fig. S12(a). Note that the system has mirror symmetry $M_x$ and $M_y$, as shown in Fig. S12(a). We take advantage of such mirror symmetries and examine only one corner and one edge, because the other three must behave in identical ways. We excite the bulk modes with a loudspeaker at the central site at three frequencies found within the five regions of strong bulk response (the gray regions in Fig. S12(b)). In the raster-maps (Fig. S12(c)), the system's mirror symmetries can be clearly identified. When the system is excited at corner I, three peaks are found at $f_I^1, f_I^2, f_I^3$, respectively (the light-gray dotted curve in Fig. S12(b)), which largely overlaps in their frequencies with the three peaks obtained at edge A (with the microphone placed several sites away from corner I). This hints that they may be due to the same set of modes. In Fig. S12(d), the field maps, which are excited by a source at corner I at $f_I^1, f_I^2, f_I^3$, shows that indeed these modes are extended bulk modes. Our predictions are therefore experimentally validated.

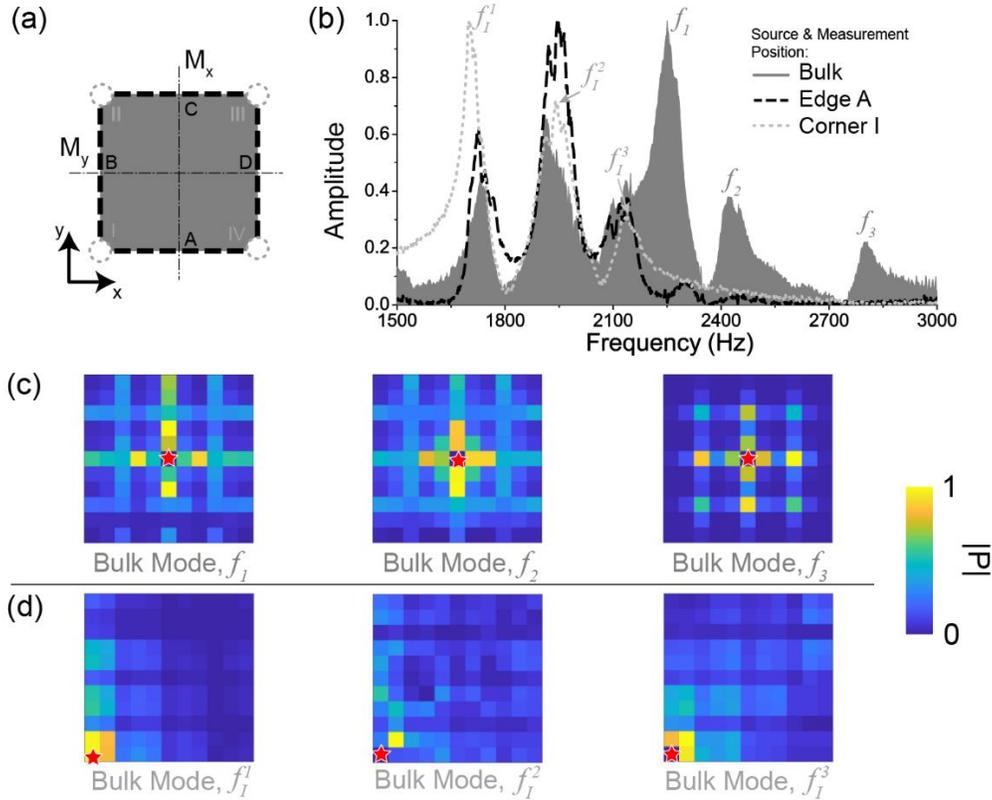

**Figure S12** Experimental results for $(\phi_x, \phi_y) = (0, 0)$. (a) A schematic drawing of the system. (b) The pressure responses with different excitation and measurement position as indicated in the legend. (c) The field maps of the bulk modes at frequencies as indicated. (d) The field maps of the bulk modes at frequencies as indicated, with the source at corner I. The red stars the (c, d) show the source positions.

**References**

[1]  Y. Yang, Z. Yang, and B. Zhang, Acoustic valley edge states in a graphene-like resonator system, J.

Appl. Phys. **123**, 091713 (2018).

[2] Y.-X. Xiao, G. Ma, Z.-Q. Zhang, and C. T. Chan, Topological Subspace-Induced Bound State in the Continuum, Phys. Rev. Lett. **118**, 166803 (2017).

[3] X. Ni, M. Weiner, A. Alù, and A. B. Khanikaev, Observation of higher-order topological acoustic states protected by generalized chiral symmetry, Nat. Mater. **18**, 113 (2019).

[4] D. R. Hofstadter, Energy levels and wave functions of Bloch electrons in rational and irrational magnetic fields, Phys. Rev. B **14**, 2239 (1976).

[5] Y. E. Kraus and O. Zilberberg, Topological Equivalence between the Fibonacci Quasicrystal and the Harper Model, Phys. Rev. Lett. **109**, 116404 (2012).

[6] A. H. MacDonald, Landau-level subband structure of electrons on a square lattice, Phys. Rev. B **28**, 6713 (1983).

[7] Y. E. Kraus, Z. Ringel, and O. Zilberberg, Four-dimensional quantum Hall effect in a two-dimensional quasicrystal, Phys. Rev. Lett. **111**, 226401 (2013).